\documentclass[a4paper,UKenglish,cleveref, autoref, thm-restate]{lipics-v2019}
\title{How to Sell Information Optimally: an Algorithmic Study} %TODO Please add

\titlerunning{How to Sell Information Optimally: an Algorithmic Study} %TODO optional, please use if title is longer than one line

\author{Yang Cai}{Yale University, USA 
}{yang.cai@yale.edu}{https://orcid.org/0000-0002-5426-1324}{Supported by a Sloan Foundation Research Fellowship and the NSF Award CCF-1942583 (CAREER).}%TODO mandatory, please use full name; only 1 author per \author macro; first two parameters are mandatory, other parameters can be empty. Please provide at least the name of the affiliation and the country. The full address is optional

\author{Grigoris Velegkas}{Yale University, USA}{grigoris.velegkas@yale.edu}{https://orcid.org/0000-0001-7148-0548
}{Partially supported by a PhD Scholarship from the Onassis Foundation and a PhD Scholarship from the Bodossaki Foundation.}

\authorrunning{Y. Cai and G. Velegkas} %TODO mandatory. First: Use abbreviated first/middle names. Second (only in severe cases): Use first author plus 'et al.'

\Copyright{Yang Cai and Grigoris Velegkas} %TODO mandatory, please use full first names. LIPIcs license is "CC-BY";  http://creativecommons.org/licenses/by/3.0/

 %TODO mandatory: Please choose ACM 2012 classifications from https://dl.acm.org/ccs/ccs_flat.cfm 

 %TODO mandatory; please add comma-separated list of keywords

\relatedversion{A full version of the paper is available at \url{https://arxiv.org/abs/2011.14570}}

\acknowledgements{We would like to thank Dirk Bergemann for helpful discussions and the anonymous reviewers for their valuable comments and suggestions.}%optional

\nolinenumbers %uncomment to disable line numbering

\hideLIPIcs  %uncomment to remove references to LIPIcs series (logo, DOI, ...), e.g. when preparing a pre-final version to be uploaded to arXiv or another public repository

\EventEditors{James R. Lee}
\EventNoEds{1}
\EventLongTitle{12th Innovations in Theoretical Computer Science Conference (ITCS 2021)}
\EventShortTitle{ITCS 2021}
\EventAcronym{ITCS}
\EventYear{2021}
\EventDate{January 6--8, 2021}
\EventLocation{Virtual Conference}
\EventLogo{}
\SeriesVolume{185}
\ArticleNo{88}
\usepackage[utf8]{inputenc}
\usepackage{amsmath,amssymb,amsthm,amsfonts,latexsym,bbm,xspace,graphicx,float,mathtools,dsfont}
\usepackage{graphicx}
\usepackage{mathtools} 
\usepackage{extarrows} 
\usepackage{bbm}
\usepackage{hyperref}
\usepackage{color}
\usepackage{times}
\usepackage{algorithm, algorithmic}
\usepackage{appendix}

\newenvironment{prevproof}[2]{\noindent {\em {Proof of {#1}~\ref{#2}:}}}{$\Box$\vskip \belowdisplayskip}

\newcommand{\A}{\mathcal{A}}

\newcommand{\N}{\ensuremath{\mathbb{N}}} % The naturals
\newcommand{\M}{\mathcal{M}}

\newcommand{\poly}{{\rm poly}}
\newcommand{\opt}{\text{OPT}}

\newcommand{\todo}[1]{{}}
\newcommand{\notshow}[1]{{}}

\newcommand{\rev}{\textsc{Rev}}

\DeclareMathOperator*{\E}{\mathbb{E}}
\DeclareMathOperator*{\argmax}{arg\,max}

\def \PP  {{\cal P}}

\def \MM  {\mathcal{M}}

\def \PP{{\mathcal{P}}}

\makeatletter
\def\Ddots{\mathinner{\mkern1mu\raise\p@
\vbox{\kern7\p@\hbox{.}}\mkern2mu
\raise4\p@\hbox{.}\mkern2mu\raise7\p@\hbox{.}\mkern1mu}}
\makeatother

\definecolor{MyGray}{rgb}{0.8,0.8,0.8}

\allowdisplaybreaks

\begin{document}

\maketitle

\begin{abstract}
%\grigorisnote{The model of Babaioff et al.~\cite{babaioff2012optimal}, Chen et al.~\cite{chen2020selling} is similar to ours so maybe not say initiate?}
    We investigate the algorithmic problem of selling information to agents who face a decision-making problem under uncertainty. We adopt the model recently proposed by Bergemann et al.~\cite{bergemann2018design}, in which information is revealed through signaling schemes called \textit{experiments}. In the single-agent setting, any mechanism can be represented as a menu of experiments. Our results show that the computational complexity of designing the revenue-optimal menu depends heavily on the way the model is specified. When all the parameters of the problem are given explicitly, we provide a polynomial time algorithm that computes the revenue-optimal menu. For cases where the model is specified with a succinct implicit description, we show that the tractability of the problem is tightly related to the efficient implementation of a \textit{Best Response Oracle}: when it can be implemented efficiently, we provide an additive FPTAS whose running time is independent of the number of actions. On the other hand, we provide a family of problems, where it is computationally intractable to construct a best response oracle, and we show that it is NP-hard to get even a constant fraction of the optimal revenue. Moreover, we investigate a generalization of the original model by Bergemann et al.~\cite{bergemann2018design} that allows multiple agents to compete for useful information. We leverage techniques developed in the study of auction design (see e.g.~\cite{CaiDW12a, AlaeiFHHM12, CaiDW12b,CaiDW13a,CaiDW13b}) to design a polynomial time algorithm that computes the revenue-optimal mechanism for selling information.
\end{abstract}

\section{Introduction}\label{sec:intro}
Decision-making heavily relies on information availability. The uneven distribution of information thus enables markets for trading information. Imagine a bank reviewing a loan application. Information about the borrower's financial status clearly influences the bank's lending decisions. In this setting, the bank already has some private knowledge about the borrower, i.e., through prior interactions, but may still be willing to pay to acquire supplemental information to guide its decision-making. Indeed, Equifax, the credit report agency, provides its business consumers, e.g., banks and credit card companies, a product called Undisclosed Debt Monitoring, that tracks negative information about individual borrowers.~\footnote{https://www.equifax.com/business/undisclosed-debt-monitoring/}

How should the information owner reveal and price the information? We adopt the model introduced by Bergemann et al.~\cite{bergemann2018design} to study this problem. Their model involves a data buyer and a data seller. The data buyer faces a decision under uncertainty, and his payoff depends on the \emph{action} he decides to take and the underlying \emph{state} of the world. Initially, the buyer only has some \emph{imperfect} knowledge about the state, i.e. a prior distribution over the possible states. This piece of information is private to the buyer. On the contrary, the data seller knows the state of the world and can sell supplemental information to the buyer. Since the buyer's willingness to pay for supplemental information is determined by the precision of his own prior belief, we refer to the buyer's prior belief as his type. How does the seller optimize her revenue if the buyer's type is assumed to be drawn from a known distribution? To screen heterogeneous buyer types, the seller offers a menu of information products. Each product has the form of a \emph{statistical experiment}, whose result reveals a signal that is correlated with the underlying state. Bergemann et al.~\cite{bergemann2018design} investigate what experiments should be included and how to price them. They obtain analytic solution of the revenue-optimal menu in two special cases: (i) the case with only two possible buyer types and (ii) the case with two states and two possible actions for the buyer to choose from. However, general characterization of the revenue-optimal menu remains elusive. In this paper, we initiate the algorithmic study of this problem and investigate the computational complexity for finding the revenue-optimal menu.

Our first result considers an explicit representation of the problem, where the input contains the buyer's type distribution and his payoff for each action and state pair.

\medskip 
%\noindent 
\hspace{0.7cm}\begin{minipage}{0.9\textwidth}
\begin{enumerate}
\item[{\bf Result I:}] We design an algorithm that computes the revenue-optimal menu in time polynomial in the number of buyer types, the number of buyer actions, and the number of underlying states. (Theorem~\ref{thm:alg explicit})
\end{enumerate}
\end{minipage}
\medskip

For many settings of interest, the model is too expensive to be specified explicitly but has a natural succinct implicit description. Consider the following motivating example. Suppose there is a traffic network $G$, and a binary state $\omega$ that indicates the level of congestion on the edges of $G$. A driver wants to go from a vertex $s$ to another vertex $t$, and his payoff is $H$ {minus} the expected travel time from $s$ to $t$ using the path he picks.~\footnote{$H$ is a sufficiently large constant so that the payoff is always nonnegative.} Of course, which path is the fastest depends on the driver's belief of the state $\omega$. Suppose Waze is offering a service that provides the driver supplemental information about the congestion. How should Waze price its service? In this setting, the driver is the data buyer and the number of actions available to him is exactly the number of $s-t$ paths,  which can be exponential in the size of the network $G$. Applying our first result is thus computationally inefficient in this setting. Our second result concerns exactly these settings with succinct implicit descriptions. We show that the key to tractability is the existence of a computationally efficient \emph{Best Response Oracle}, that is, an oracle that accepts a distribution over the underlying state as input and outputs an action with the highest expected payoff. Clearly, in the example above, it is straightforward to construct a computationally efficient best response oracle -- simply assign the expected congestion on each edge as its length and run any shortest path algorithm on the graph $G$. We show that finding the revenue-optimal menu is tractable as long as there exists a best response oracle.%, an oracle  In this section, we show an algorithm that computes an up-to-$\varepsilon$ optimal menu in this setting efficiently. The key feature of this setting that allows us to design such an algorithm is that the best response oracle can be implemented efficiently. Namely, given any posterior belief over the states, one can run any shortest path algorithm to compute the best path in time polynomial in the natural description size of $G$. Indeed, we prove a more general result, that is, given access to a best response oracle, our algorithm can

\medskip 
%\noindent
\hspace{0.7cm}\begin{minipage}{0.9\textwidth}
\begin{enumerate}
\item[{\bf Result II:}] For any setting with a constant number of possible underlying states, we design an FPTAS to compute an \emph{up-to-$\varepsilon$ optimal menu}, i.e., a menu whose revenue is at most $\varepsilon$ less than the optimum, given access to a best response oracle.~\footnote{Our algorithm runs in time polynomial in the total number of buyer types and $1/\varepsilon$.} (Theorem~\ref{thm:lp gives approx})
\end{enumerate}
\end{minipage}
\medskip

%\noindent
Without the best response oracle, we show that it is NP-hard to even find a constant factor approximation to the optimal revenue  for a family of succinctly describable instances with only two underlying states and one buyer type. (Theorem~\ref{thm:sat reduction})

We also investigate extensions of the basic model studied in Bergemann et al.~\cite{bergemann2018design}. First, we consider the setting where multiple data buyers are competing with each other to receive an informative signal from the data seller. We obtain the following generalization of Result I.

\medskip
%\noindent
\hspace{0.7cm}\begin{minipage}{0.9\textwidth}
\begin{enumerate}
\item[{\bf Result III:}] We design an algorithm that computes the revenue-optimal mechanism in time polynomial in the \emph{total number of buyer types}, the number of buyer actions, and the number of underlying states. (Theorem~\ref{thm:mutli-agent})
\end{enumerate}
\end{minipage}
\medskip

%\noindent 
Note that the straightforward generalization of Result I only gives an algorithm that runs in time polynomial in the \emph{total number of buyer type profiles}, which is exponential in the number of buyers. Our algorithm in Result III runs in time polynomial in the \emph{total number of buyer types}, which is the description size for specifying each buyer's type distribution.

In {Section}~\ref{sec:further extensions}, we discuss another natural extension. We can treat a buyer's payoffs for taking actions in various states as the buyer's private information and may be different across buyers. In other words, a buyer type is no longer just the buyer's prior belief of the underlying state but also his payoff function. We show that all our results (c.f. Result I, II, and III) can be easily extended to handle this case, and the modification is summarized in {Section}~\ref{sec:further extensions}.

\vspace{-.15in}
\paragraph*{Our Approach:} In the explicit model, we first show that the revenue-optimal menu can be captured by a LP with polynomially many decision variables but exponentially many constraints. {We leverage a technique introduced by Chen et al.~\cite{chen2020selling} to transform it to an equivalent LP with polynomially many constraints.} %Luckily, we show that there is a polynomial time algorithm that checks all the constraints, which allows us to apply the ellipsoid method to solve the LP.
In the implicit model with a best response oracle, the main difficulty is that the optimal menu may contain experiments that use an arbitrary number of signals, and as a result the menu cannot even be represented in polynomial time. To overcome this difficulty, we first argue that there always exists an up-to-$\varepsilon$ optimal menu that only contains experiments that use a small number of signals, then apply {an algorithm similar to the one} in the explicit model to find such a menu. For the multi-agent setting, one can use a LP similar to the one in the explicit model to capture the optimal mechanism. The issue with this approach is that both of the number of variables and constraints are exponential in the number of agents, which is too large to solve. Our solution is inspired by an approach used to computing the revenue-optimal multi-item auctions~\cite{CaiDW12a,AlaeiFHHM12,CaiDW12b,CaiDW13a,CaiDW13b}. The key idea is to first represent mechanisms in a succinct way known as the ``reduced forms'' and use a LP to search for the revenue-optimal reduced form. However, as a reduced form is only a succinct description, given a reduced form, it is not obvious what mechanism it corresponds to. Indeed, it is not even clear whether it corresponds to any mechanism. The main technical barrier we overcome is to design efficient algorithms to (i) check the feasibility of a reduced form and (ii) to implement a feasible reduced form as an actual mechanism.

\subsection{Related Work}
\paragraph*{Relationship with Monopoly Pricing.} A well-studied problem from mechanism design, the \emph{monopoly pricing problem}, bears some resemblance to our problem. The monopoly pricing problem asks what is the revenue-optimal mechanism to sell one or more items to a buyer, whose valuation/willingness to pay for the items is drawn from a known distribution. In single-dimensional settings, the problem is completely resolved by~\cite{Myerson81, riley1983optimal}. In multi-dimensional settings, complete characterizations are known only in several special cases~\cite{GiannakopoulosK14,GiannakopoulosK15,HaghpanahH14,daskalakis2017strong, fiat2016fedex, devanur2017optimal, devanur2020optimal}, but simple and approximately optimal mechanisms~\cite{ChawlaHK07,ChawlaHMS10,HartN12,BabaioffILW14,LiY13,Yao15,CaiDW16,CaiZ17}, as well as algorithmic characterizations~\cite{CaiDW12a,AlaeiFHHM12,CaiDW12b,CaiDW13a,CaiDW13b} have been discovered in fairly general settings. The nature of information goods adds an extra layer of difficulty to the pricing problem. The value of information is determined by how much such information can improve the quality of decision-making. Buyers with different beliefs do not simply have different values for different experiments, but they may even disagree on their ranking. This richness in buyer valuations does not happen in single-dimensional monopoly pricing, but already exists in single-dimensional information pricing, where there are only two possible underlying states. \footnote{In this case, the buyer's prior belief can be represented using a single real number in $[0,1]$.} As a result, the optimal menu in information pricing has a more complex structure. {For example, Bergemann et al.~\cite{bergemann2018design} showed that even in the two-state case, the seller sometimes needs to use randomized experiments to maximize her revenue.} In this work, we show that despite the new challenges of selling information products, some of the algorithmic ideas from monopoly pricing are still useful.

\vspace{-.1in}
\paragraph*{Relationship with Information Design.} Similar to our problem, the designer constructs a signaling scheme that reveals partial information about the underlying state to influence the action of the agents in information design~\cite{KamenicaG11}. There has been growing interest in the algorithmic study of information design~\cite{DughmiIR14,dughmi2014hardness,DughmiX16,cheng2015mixture,DaskalakisPT16,cai2020third}. The fundamental difference between our setting and information design is that, in our setting, the buyer's action does not have direct effect on the seller's utility. The seller only derives utility from the monetary transfers received from the buyer.

\vspace{-.1in}
\paragraph*{Relationship with other Information-Selling Models.}
The work that is most related to the problem we are studying is by Babaoiff et al.~\cite{babaioff2012optimal}. Similar to us, they consider a seller who knows the state of the world $\omega$ and wants to sell information to a buyer whose prior is drawn from some distribution $\Theta$. However, there is a subtle but crucial difference between our models. In their work, the seller's information disclosure strategy and the price are allowed to be dependent on the realized state of the world $\omega$. By contrast, our model requires the seller to commit to a mechanism before the realization of the state $\omega$. Their main results state that (i) when $\omega$ and the buyer's type are independently distributed, revelation principle holds, i.e., a single-round interaction suffices; and (ii) when $\omega$ and the buyer's type are correlated, a full surplus extraction mechanism, similar to Cr\'{e}mer and McLean~\cite{CM88}, exists and can be computed efficiently. These results differ quite substantially from the structural results for the model we study here~\cite{bergemann2018design}. For instance, full surplus extraction is in general impossible in our model. %Since the buyer's prior could be correlated with $\omega$, the optimal mechanism may need more than one round of interaction between the buyer and the seller. They identify conditions under which the optimal mechanism consists of just one interaction between the two entities and show an efficient algorithm that computes the optimal mechanism using the ellipsoid method. Moreover, they provide  inapproximability results of the optimal revenue by simple mechanisms in the general case. 
In a recent work, Chen et al.~\cite{chen2020selling} extends the results by Babaioff et al.~\cite{babaioff2012optimal} to the setting where the buyer is budget-constrained, and improves upon some of the algorithmic results. %They also identify settings in which the optimal revenue is achieved by one interaction between the buyer and the seller.
%They also improve upon the previous algorithmic results showing that the optimal mechanism can be computed efficiently using a polynomial size LP rather using the ellipsoid algorithm.

\section{Preliminaries}

\paragraph*{Model and Notation.} The data buyer faces a decision problem under uncertainty. The state of the world $\omega$  is drawn from a state space $\Omega$. The buyer chooses an action $a$ from an action space $A$. Throughout the paper, we use $m$ to denote the size of $A$. The buyer's \textbf{ex-post utility} for choosing action $a$ under state $\omega$ is defined to be  $u_{\omega, a}$ and is assumed to lie in $[0,1]$. The buyer has some prior information about the state of the world which is denoted by $\theta$ and comes from a set $\Theta \subseteq \Delta \Omega$. We call $\theta$ the \textit{type} of the buyer, and use $\theta_{\omega}$ to denote the probability that the buyer assigns to the event that the state of the world is $\omega$. The type of the buyer is distributed according to $F$. Apart from the buyer, there is also a \textit{seller} who observes the state of the world and is willing to sell supplemental information to the buyer. We refer to the buyer as he and to the seller as she.

%The objective of the seller is to maximize her revenue and does not depend on the action the buyer takes. 
\vspace{-.1in}
\paragraph*{Experiment.} The seller provides supplemental information to the buyer via a \textit{signaling scheme} which we call \textit{experiment}. A signaling scheme is a commitment to $|\Omega|$ probability distributions over a set of different signals $S$, such that when the state of the world is realized, the seller draws a signal from the corresponding distribution and sends it to the buyer.
We denote such an experiment by $E = (S, \pi(E))$, where  $\pi(E): \Omega \rightarrow \Delta S$ denotes the distributions that experiment $E$ is using. We denote the probability that experiment $E$ sends signal $s_k$ when the state of the world is $\omega$ by $\pi_{\omega, k}(E) = \Pr[s_k| \omega]$. {In this work, it is useful to think {of the experiment $\pi(E)$ as a matrix whose rows are indexed by the states of the world and columns are indexed by the signals.}} A \textit{menu} of experiments is a collection $\MM = \{(E, t(E))\}$, where $t(E) \in [0,1]$ is the payment the buyer has to make when he purchases experiment $E$. In the multi-agent setting, a menu is insufficient to generate the optimal revenue, and our goal there is to compute the revenue-optimal \emph{mechanism}. 
In the single-agent setting, the interaction between the seller and the buyer works as follows:
\begin{enumerate}
    \item The seller posts a menu $\MM$.
    \item The state of the world $\omega$ and the type of the buyer $\theta$ are realized.
    \item The buyer chooses some experiment $E$ from the menu based on his type and pays $t(E)$.
    \item The seller sends the buyer a signal $s$ that is drawn {from $\pi_{\omega, \cdot}(E)$}.
    \item The buyer chooses an action  $a$, based on his original belief $\theta$ and the signal $s$, and receives utility $u_{\omega, a}$.
\end{enumerate}

\vspace{-.1in}
\paragraph*{The Value of an Experiment.} To understand the behavior of the buyer, we first explain how the buyer evaluates an experiment. %We denote by $E(\theta)$ the experiment that type $\theta$ prefers.
We first explain how the buyer would act if the only information available to him was his type $\theta$. Since the buyer picks an action that maximizes his expected utility, his best move without receiving any additional information from the seller is $a(\theta) ={ \argmax_{a} \sum_\omega \theta_\omega u_{\omega, a}}$ and his  \textbf{base utility} following that move is {$u(\theta) = \max_{a} \sum_\omega \theta_\omega u_{\omega, a}$.} If he receives extra information from the seller, he updates his beliefs and may choose a new action that induces higher expected value based on his posterior distribution over the states. After receiving signal $s_k$ from experiment $E$ his belief about the state of the world is
$$ \Pr[\omega| s_k, \theta] = \frac{\theta_\omega \pi_{\omega, k}(E)}{\sum_{\omega'\in \Omega}\theta_{\omega'}\pi_{\omega', k}(E)}.
$$
Hence, the best action is
$$    a(s_k|\theta) \in \argmax_a \sum_\omega \left(\frac{\theta_\omega \pi_{\omega, k}(E)}{\sum_{\omega'\in \Omega} \theta_{\omega'} \pi_{\omega', k}(E)}\right)u_{\omega, a},$$
which yields \textbf{conditional expected utility}
$$u(s_k|\theta) := \max_a \sum_\omega \left(\frac{\theta_\omega \pi_{\omega, k}(E)}{\sum_{\omega'} \theta_{\omega'} \pi_{\omega', k}(E)}\right)u_{\omega, a}.$$
{When it is clear from the context, we might drop $\theta$ in the previous expressions.} Notice that after computing his posterior, the buyer's conditional expected utility is linear in the actions so we can assume w.l.o.g. that he picks a single action and not a distribution over actions. Taking the expectation over the signal he will receive, we denote the \textbf{value of the experiment} $E$ for type $\theta$ to be 
$$ V_{\theta}(E) = \sum_{s_k\in S} \max_a \left\{ \sum_{\omega} \theta_{\omega} \pi_{\omega, k}(E) u_{\omega, a} \right\}.$$

We assume that the buyer is quasilinear. We denote the \textbf{expected net utility} of type $\theta$ for experiment $E$ as $V_\theta(E)-t(E)$.

Notice that two different types $\theta, \theta'$ may have different favorite actions under the same signal. As a result, $V_\theta(E)$ is not a linear function over $\theta$ for a fixed experiment $E$ even when the type $\theta$ is single-dimensional, i.e., $|\Omega|=2$. 
Consider the following example.\\
\begin{example}
There are two states of the world $\Omega = \{\omega_1, \omega_2\}$, two possible actions $A = \{a_1, a_2\}$ and we have a \emph{matching action-state} environment, i.e. the ex-post utilities are as follows
$$
U = \begin{pmatrix}
1 & 0\\
0 & 1
\end{pmatrix}.
$$
Consider the following experiment $E$ that sends two different signals $s_1, s_2$ with
$$
E = \begin{pmatrix}
0.7 & 0.3\\
0.3 & 0.7
\end{pmatrix}.
$$
For a buyer whose prior is $\theta_{\omega_1} = \theta, \theta_{\omega_2} = 1-\theta$, after receiving $s_1, s_2$ the posteriors of the buyer are $\left(\frac{0.7 \theta}{0.4\theta + 0.3}, \frac{0.3(1-\theta)}{0.4\theta + 0.3} \right)$ and $\left(\frac{0.3\theta}{0.7-0.4\theta}, \frac{0.7(1-\theta)}{0.7-0.4\theta}\right)$ respectively. Hence, we see that when $\theta \leq 0.3$ the buyer will take action $a_2$ after receiving either $s_1$ or $s_2$. Similarly, if $0.3 < \theta \leq 0.7$ he will take action $a_1$ (or $a_2$) after receiving $s_1$ (or $s_2$). Finally, when $\theta > 0.7$ the buyer will take action $a_1$ no matter which signal he receives. This leads to the following value function for $E$
\[   
V_\theta(E) = 
     \begin{cases}
       1-\theta &\quad\text{if } 0 \leq \theta \leq 0.3 \\
       0.7 &\quad\text{if } 0.3 < \theta \leq 0.7 \\
       \theta &\quad\text{if } 0.7 \leq \theta \leq 1 \\
     \end{cases}
\]
\end{example}
This is in sharp contrast to standard single-dimensional auction design settings, where the agent's value is always a linear function over the type when the allocation is fixed.~\footnote{Indeed, this linearity even holds for quite general multi-dimensional settings in auction design, for example, when the buyer has additive valuations.}  %, since the posterior distributions that the signals of the experiment induce might be vary significantly.

\vspace{-.15in}
\paragraph*{IC and IR Menu.}
The buyer chooses the experiment that gives him the highest net utility, and we slightly abuse notation and denote
 by $E(\theta)$ the experiment of the menu $\MM$ that type $\theta$ prefers
$$E(\theta) = \argmax_E V_{\theta}(E) - t(E).$$ %Notice \grigorisnote{a slight difference with the auction setting}. 
 If all experiments give him expected net utility smaller than his base utility $u(\theta)$, he will not purchase any experiment. For convenience, we assume that there is a null experiment that provides no information offered at price $0$, so now without loss of generality every type takes an experiment from the menu. %We use $E(\theta)$ and $t(\theta)$ to denote the experiment and price that type $\theta$ selects.
We sometimes abuse notation to call a menu \textit{Incentive Compatible} (IC) and \textit{Individually Rational} (IR). By that, we mean that every buyer type $\theta$ selects the experiment $(E(\theta),t(\theta))$ that maximizes his expected {net utility}. %if  every type $\theta \in \Theta$  prefers the selected experiment $\left(E(\theta),t(\theta)\right)$ to any other pair $(E, t(E)) \in \MM$ and his utility derived from this experiment is greater than his initial utility. 
Formally, these constraints are captured by the following two sets of inequalities:
\begin{align*}
    V_\theta(E(\theta)) - t(\theta) &\geq V_{\theta}(E(\theta')) -t(\theta'), ~~~\forall \theta, \theta' \in \Theta, \\
    V_\theta(E(\theta)) - t(\theta) &\geq u(\theta), ~~~\forall \theta \in \Theta.
\end{align*}
Once we have fixed the parameters of the model, we denote by $\rev(\MM)$ the revenue that menu $\MM$ generates and by $\opt$ the optimal revenue in this setting. 

%\paragraph*{Notation for the multi-agent setting:}  We now introduce the additional notation we use in the multi-agent setting. We use $\Theta^i$ to denote the type space of buyer $i$ and $F^i(\theta^i)$ to denote the probability that buyer $i$'s type is $\theta^i$. We use $\Theta$ to denote the set of all type profiles and $F(\theta)$ to denote $\times_{i\in[n]} F^i(\theta^i)$. We assume the action space $A$ is the same for each buyer $i$, but the payoff $u^i_{\omega,a}$ for choosing action $a$ under state $\omega$ may be different for different buyers. %We consider the \emph{explicit model}, that is, for each buyer $i$, both $F^i$ and the payoff matrix $U^i=\{u^i_{\omega,a}\}_{\omega\in \Omega, a\in A}$ are given as input.  We use $u^i(\theta^i)$ to denote the payoff of buyer $i$ for choosing the best action under distribution $\theta^i$.

\section{Optimal Menu for a Single Agent}\label{sec:single-agent} In this section, we consider the problem of computing the optimal menu for a single agent. As we will show, the computational complexity of the problem is tightly related to the way that the problem is specified. We consider three different models.
\begin{itemize}
    \item \textbf{Explicit model:} the distribution $F$ and the ex-post utility matrix $U$ are given explicitly in the input.
    \item \textbf{Implicit model with a best response (BR) oracle:} the distribution $F$ is given along with a best response oracle. The best response oracle accepts a distribution over the states as input and outputs the action that generates the highest expected utility w.r.t. the input distribution.
    \item \textbf{Implicit model with succinct description:} the distribution $F$ is given along with the description of a Turing Machine that computes the ex-post utility of any pair of state and action in time polynomial in the description size. %a succinct representation of the payoff matrix $U$. 
    %Importantly, $A$ exponentially large in this representation.}
\end{itemize}

We summarize our results with respect to the three different models: (i) we show that computing the optimal menu in the explicit model is capture{d} by  a polynomial size LP; % an \emph{exponential size LP} which can be solved efficiently using the ellipsoid method;
(ii) even though the number of actions in the implicit model with BR oracle may be arbitrarily large, we provide an FPTAS to find an up-to-$\varepsilon$ optimal menu for settings with a constant number of underlying states; (iii) we construct a succinctly representable instance of the problem such that computing a constant factor approximation or an up-to-$\varepsilon$ optimal menu is NP-hard.

\subsection{Explicit Model}\label{sec:explicit model}
In this section, we discuss the basic setting where the model is explicitly given.  First, we state a structural result of the optimal menu that allows us to restrict the number of signals.

\begin{definition}[Responsive Experiment]\label{def:responsive experiment}
A buyer type $\theta$ is \emph{responsive} to an experiment $E$ if every signal $s$ of $E$ leads $\theta$ to a different optimal choice of action and, in particular $a(s_k\mid \theta) = a_k$ for all $s_k\in S$. 
\end{definition}

Note that if an experiment $E$ is responsive to any buyer type, $E$ has exactly $m$ signals. An intuitive way to think about Definition~\ref{def:responsive experiment} is that signal $s_i$ of $E$ recommends the buyer to take a action $a_i$, and type $\theta$ is responsive to $E$ if $\theta$ always follow the recommendation. Importantly, a different type $\theta'$ may not be responsive to  experiment $E$ and does not follow the recommendations. We now state the structural result from~\cite{bergemann2018design} that states that it is without loss of generality to consider a menu where each buyer type purchases a responsive experiment.

\begin{lemma}[Adapted from Proposition 1 from~\cite{bergemann2018design}]\label{lem:responsive menu}
   A menu $\M$ is \emph{responsive} if for every buyer type $\theta$, it chooses a responsive experiment $E(\theta)$ from the menu. We define the outcome of a menu as the joint distribution of states, actions, and monetary transfers resulting from every buyer type’s optimal choice of experiment and subsequent choice of action.  The outcome of every menu can be attained by a responsive menu.
\end{lemma}

The proof of Lemma~\ref{lem:responsive menu} is based on a revelation-principle type of argument. More precisely, assume that some type $\theta$ prefers an experiment $E(\theta)$ for which he is not responsive. Then, we can merge all the signals $s_{i_1}, \ldots, s_{i_\ell}$ of $E(\theta)$ that lead the type to take the same action $a_k$ and just send the merged signal as the new $k$-th signal $s_k$. Equipped with the structural result, we are ready to show how to capture the design of the optimal menu as a LP. Since there exists an optimal menu that is responsive,  every experiment in this menu has at most $m$ signals. For every buyer type $\theta$, we use $\pi(\theta)$ to denote the experiment type $\theta$ purchases, where $\pi_{\omega,i}(\theta)$ is the probability to send signal $s_i$ when the state is $\omega$. $\{\pi_{\omega,i}(\theta)\}_{\omega\in \Omega, i\in [m], \theta\in \Theta}$ is the first set of variables. We also have the prices for each experiment $\{t(\theta)\}_{\theta\in \Theta}$ as variables. The main difference between selling experiments and selling items is that different types may interpret the same experiment differently. More specifically, for a responsive experiment $(\pi(\theta),t(\theta))$, the optimal choice of action for type $\theta$ after receiving signal $s_i$ is simply action $a_i$; while for a different type $\theta'$, the best action after receiving signal $s_i$ can be a completely different action $a_j$ due to a different induced posterior. As our LP is designed to compute the optimal and responsive menu, we need to guarantee that for any type $\theta$, purchasing experiment $(\pi(\theta,t(\theta))$ and following the recommendation is better than purchasing any experiment $(\pi(\theta'),t(\theta'))$ for a different type $\theta'$ and subsequently choosing the optimal action based on the induced posterior. We remark that the straightforward formulation of the previous constraints requires one to consider all the possible mappings from signals to actions, resulting in a LP that has exponentially many constraints in $m$. Although this huge LP can be solved using the ellipsoid algorithm, we use a technique inspired by Chen et al.~\cite{chen2020selling} that allows us to formulate it using only polynomially many constraints in $m$.

See Figure~\ref{fig:alg explicit} for our LP. Besides the variables representing the experiments and prices, we introduce a new set of variables $\left\{z_i(\theta,\theta')\right\}_{i\in [m], \theta, \theta'\in \Theta}$. $z_i(\theta, \theta')$ serves as an upper bound of the conditional expected utility that $\theta$ gets when he considers misreporting as $\theta'$ and receives signal $s_i$, which is guaranteed by the second set of constraints. The first set of constraints make sure that for any type $\theta$, the net utility of purchasing experiment $E(\theta)$ and following its recommendations is no worse than the utility of purchasing any other experiment and subsequently choosing the best action under each signal. Note that we allow $\theta'$ to be the same $\theta$ in the first set of constraints, which guarantees that the menu is responsive. The third set of constraints guarantees that purchasing experiment $(\pi(\theta),t(\theta))$ is no worse than $\theta$'s base utility. We refer to these constraints as the individual rationality (IR) constraints. The last two constraints guarantee that $\pi(\theta)$ is indeed an experiment. %The third set of constraints make sure the menu is IR, that is, the net utility of experiment $E(\theta)$ is better than the base utility of $\theta$.

%The first set of constraints provide \grigorisnote{the previous} guarantee. {It is not very hard to see that because we are maximizing the revenue, $z_i(\theta, \theta')$ better not be greater that the expected payoff after receiving $s_i$. Notice that if we choose $\theta'$ to be $\theta$ we have that $\sum_{i \in [m]} z_i(\theta, \theta)$ is at most as big as the expected payoff from following the recommendation. This combined with the third set of constraints for $z_i(\theta, \theta)$ guarantees that $\theta$ is obedient to the recommendations of the experiment.} The LHS is the payoff when type $\theta$ follows the recommendation of $(\pi(\theta),t(\theta))$, and the RHS is the payoff when type $\theta$ purchases experiment $(\pi(\theta'),t(\theta'))$ and chooses action $a_{\sigma(i)}$ when the experiment sends signal $s_i$. As we allow $\sigma$ to be any mapping from $[m]$ to $[m]$, the maximum value of the RHS is the largest net utility type $\theta$ can derive from buying experiment $(\pi(\theta'),t(\theta'))$. Note that if we choose $\theta'$ to be $\theta$, the first set of constraints guarantees that $\theta$ is responsive to experiment $(\pi(\theta),t(\theta))$. We refer to the first set of constraints the incentive compatibility (IC) constraints. %The second set are the IC constraints, which imply that every type should simply report her true value. They are exponentially many, since whenever she gets a signal that was not intended for her, she has to figure out the best action that corresponds to that posterior. 

\begin{figure}[h!]
\colorbox{MyGray}{
\begin{minipage}{0.97\textwidth} {
\noindent\textbf{Variables:}
\begin{itemize}
\item $\{\pi_{\omega,i}(\theta)\}_{\omega\in \Omega, i\in [m], \theta\in \Theta}$, denoting the experiments in the menu.
\item $\{t(\theta)\}_{\theta\in \Theta}$, denoting the prices of the experiments. 
\item $\{z_i(\theta, \theta')\}_{i \in [m], \theta, \theta' \in \Theta}$, helper variables. $z_i(\theta, \theta')$ represents an upper bound of the conditional expected utility of signal $s_i$ from experiment $E(\theta')$ for type $\theta$.
\end{itemize}
\textbf{Linear Program:}
\begin{equation*}
\begin{array}{ll@{}ll}
\text{max}  & \displaystyle\sum\limits_{\theta\in \Theta} F(\theta)t(\theta)&\\
% \text{subject to} &\displaystyle\sum\limits_{\omega}\theta_{\omega} \pi_{\omega, i}(\theta)u_{\omega, a_i} \geq 
%                     \displaystyle\sum\limits_{\omega}\theta_{\omega} \pi_{\omega, i}(\theta)u_{\omega, a_j},  & \forall \theta, \forall i, j \in [m] ~~~(\text{OB})\\
    \text{s.t.}            &  \displaystyle\sum\limits_{i\in [m]}\displaystyle\sum\limits_{\omega\in \Omega} \theta_{\omega} \pi_{\omega, i}(\theta)u_{\omega, a_i} - t(\theta) \geq
                    \displaystyle\sum\limits_{i\in [m]}z_i(\theta, \theta') - t(\theta'),~~~ 
                &   \forall \theta, \theta'\in \Theta ~~~(\text{IC})\\
                & z_i(\theta, \theta') \geq \displaystyle\sum_\omega \theta_\omega \pi_{\omega, i}(\theta') u_{\omega, a_j}, &\forall \theta, \theta' \in \Theta, \forall  i, j \in [m]\\
                & \displaystyle\sum\limits_{i\in [m]}\displaystyle\sum\limits_{\omega\in \Omega} \theta_{\omega} \pi_{\omega, i}(\theta)u_{\omega, a_i} - t(\theta) \geq u(\theta),
                & \forall \theta\in \Theta ~~~(\text{IR})\\  
                & \displaystyle\sum\limits_{i\in [m]} \pi_{\omega, i} (\theta) = 1, & \forall \theta\in \Theta, \omega\in \Omega \\
                & \pi_{\omega, i}(\theta) \geq 0, & \forall \theta\in \Theta, \forall \omega\in \Omega, \forall i\in [m]
\end{array}
\end{equation*}}
\end{minipage}} \caption{A linear program to find the revenue-optimal menu in the explicit model.}\label{fig:alg explicit}
\end{figure}

\begin{theorem}\label{thm:alg explicit}
The LP in Figure~\ref{fig:alg explicit} can be solved in time polynomial in $m=|A|$, $|\Omega|$, and $|\Theta|$, and its optimal solution is the revenue-optimal menu.
\end{theorem}
\begin{prevproof}{Theorem}{thm:alg explicit}
   The number of constraints and the number of variables of the LP in Figure~\ref{fig:alg explicit} is polynomial in $m, |\Omega|, |\Theta|$, so it is clear that it can be solved in polynomial time. We now argue that its solution is indeed the optimal responsive menu, hence the optimal menu as guaranteed by Lemma~\ref{lem:responsive menu}. Observe that every responsive menu corresponds to a feasible solution of the LP. This is because if $\{(\pi(\theta), t(\theta))\}_{\theta \in \Theta}$ is a responsive menu, then we can satisfy all the constraints by setting $z_i(\theta, \theta') = \max_{a_j} \sum_\omega \theta_\omega \pi_{\omega, i}(\theta') u_{\omega, a_j}$. Conversely, note that every feasible solution of the LP induces a responsive menu that simply chooses the experiment $E(\theta)$ as $\left(\pi(\theta), t(\theta)\right)$ for each type $\theta \in \Theta$. This is because (i) the first and second sets of constraints guarantee that  $E(\theta)$ is a responsive experiment for $\theta$ (this follows from setting $\theta'$ to be $\theta$ in both the first and second set of constraints) and $V_\theta(E(\theta)) - t(\theta) \geq V_\theta(E(\theta')) - t(\theta')$ ; (ii) the IR constraints ensure that $V(E(\theta)) - t(\theta) \geq u(\theta)$. Hence, the optimal solution corresponds to the revenue-optimal menu.
   \end{prevproof}

%    As the feasible set of the LP in Figure~\ref{fig:alg explicit} is exactly the set of all responsive menus, the optimal solution of the LP is the responsive menu with the highest revenue. Due to Lemma~\ref{lem:responsive menu}, there exists a revenue-optimal menu that is also responsive, so the LP's optimal solution is a revenue-optimal menu.
    
 %   Next, we argue that the LP can be solved in polynomial time. Note that the LP has exponentially many constraints, as there are $m^{m}$ different mappings $\sigma$. We solve the LP by first constructing a polynomial time separation oracle for the variables {and} then applying the ellipsoid method. Since there are only polynomially many IR constraints as well as the last two sets of constraints, we can check them one by one. There are indeed exponentially many IC constraints, but note that for every pair of types $\theta$ and $\theta'$, the maximum value of the RHS among all corresponding IC constraints is achieved when type $\theta$ chooses the best action $a$ for every signal from experiment $\pi(\theta')$, which can be computed by going through all available actions under every signal. In other words, for every pair of types $\theta$ and $\theta'$, the tightest constraint can be identified in time $O(|\Omega|m^2)$. Hence, we can check all IC constraints in time $O(|\Theta|^2|\Omega|m^2)$.

\subsection{Implicit Model with a Best Response Oracle}
In this section, we study the case where the model is provided implicitly. We prove that, given access to a best response oracle, our algorithm can compute an up-to-$\varepsilon$ optimal menu in time {$\poly\left(|\Theta|, \frac{|\Omega|^{|\Omega|^2}}{\varepsilon^{|\Omega|^2 + |\Omega|}}\right)$}, which is independent of the number of actions. 

We first prove a structural result which shows that no matter how large the actual action set $A$ may be, there is some $A' \subseteq A$ such that a menu $\MM$ that recommends actions only from $A'$ has negligible revenue loss compared to $\opt$.  Moreover, $|A'|$ depends only on $|\Omega|$ and the additive approximation error. We also  show that these sets can be computed efficiently.

\begin{theorem}\label{thm:num signals approx}
For any constant $\varepsilon > 0$ , given access to a BR oracle we can compute for each type $\theta \in \Theta$ a set of actions $A_\theta$ by querying $O\left( \frac{|\Omega|^{|\Omega|^2}}{\varepsilon^{|\Omega|^2 + |\Omega|}}\right)$ times the BR oracle, so that there exists an IC and IR menu $\MM$, whose experiments all contain no more than  $O\left( \frac{|\Omega|^{|\Omega|^2}}{\varepsilon^{|\Omega|^2 + |\Omega|}}\right)$ many signals. Moreover, every type $\theta$ only uses actions from $A_\theta$ upon receiving any of these signals generated by his experiment. Finally, the revenue of $\M$ is at least $\opt - O(\sqrt{\varepsilon})$.
\end{theorem}

We present some lemmas that are used in the proof of Theorem~\ref{thm:num signals approx}. Firstly, we show that given a menu $\MM$ and  $\varepsilon$, we can create a menu $\MM'$ with the following two properties: (i) the number of signals that $\MM'$ uses depends only on $|\Omega|$ and $\varepsilon$, and every type $\theta$ values the new experiment he gets at most $O(\varepsilon)$ less than his original experiment. The proof is based on the idea that merging signals of an experiment $E$ that are close does not decrease the value of the experiment by much.

\begin{lemma}\label{lem:merge signals}
Let $\varepsilon > 0$ be some given constant and let $E = \{E(\theta)\}_{\theta \in \Theta}$ be a set of experiments, where each $E(\theta)$ uses an arbitrary number of signals. Then, we can create a set of experiments $E' = \{E'(\theta)\}_{\theta \in \Theta}$ that uses $O((\frac{|\Omega|}{\varepsilon})^{|\Omega|})$ signals per player such that $V_{\theta}(E'(\theta)) \geq V_{\theta}(E(\theta)) - 2\varepsilon$.
\end{lemma}

\begin{prevproof}{Lemma}{lem:merge signals}
Fix some $\theta \in \Theta$ and let $E(\theta)$ be the experiment that is offered to this type. 
%We start by showing that we create a new experiment $E'_\theta$, which uses $|K| = \Theta((\frac{|\Omega|}{\varepsilon})^{|\Omega|})$ 
%different signals and the value $V_\theta(E')$ for this experiment is at least $V_\theta(E) - 2|\Omega|\varepsilon$.
For any signal $s_i$ that the experiment $E({\theta})$ sends we define $\hat{\pi}_{\omega, i}(\theta) =  \frac{\pi_{\omega, i}(\theta)}{\sum_{\omega'}\pi_{\omega', i}(\theta)}$ to be the normalized probability of sending signals $s_i$ in state $\omega$. We partition all the vectors $\hat{\pi}_{\cdot, i}$ according to the following procedure. Let $\tilde{\pi}_{\cdot, i}$ be the vector that is created by rounding the entries of  $\hat{\pi}_{\cdot, i}$ to multiples of $\varepsilon/|\Omega|$. We put $\hat{\pi}_{\cdot, i}, \hat{\pi}_{\cdot, j}$ in the same set of the partition if they round to the same vector, i.e. $\tilde{\pi}_{\cdot, i} = \tilde{\pi}_{\cdot, j}$. Notice that this forms an actual partition, all the $\hat{\pi}_{\cdot, i}, \hat{\pi}_{\cdot, j}$ that are in the same set have $|\hat{\pi}_{\omega, i} - \hat{\pi}_{\omega, j}| \leq \frac{\varepsilon}{|\Omega|}$ and there are at most $O\left((|\Omega|/\varepsilon)^{|\Omega|} \right)$ different sets. We now describe a merging procedure of the signals that will create $E'(\theta)$ and guarantees that the number of signals that $E'(\theta)$ uses is bounded by a number that is independent of the signals in $E(\theta)$. Moreover, $V_\theta(E'(\theta))$ has a negligible decrease compared to $V_\theta(E(\theta))$.
Assume that $\hat{\pi}_{\cdot, i}, \hat{\pi}_{\cdot, j}$ are in the same set of the partition.
Without loss of generality, let $\sum_{\omega'}\pi_{\omega', i}(\theta) \leq \sum_{\omega'}\pi_{\omega', j}(\theta)$. Then, we have that
\begin{align*}
    &\sum_{\omega'} \theta_{\omega'} \pi_{\omega', j}(\theta) u_{\omega', a_j} \geq 
    \sum_{\omega'} \theta_{\omega'} \pi_{\omega', j}(\theta) u_{\omega', a_i}\\
     \implies &\frac{\sum_{\omega'} \theta_{\omega'} \pi_{\omega', j}(\theta) u_{\omega', a_j}}{\sum_{\omega}\pi_{\omega, j}(\theta)} \geq
    \frac{\sum_{\omega'} \theta_{\omega'} \pi_{\omega', j}(\theta) u_{\omega', a_i}}{\sum_{\omega}\pi_{\omega, j}(\theta)} \\
    \implies & \frac{\sum_{\omega'} \theta_{\omega'} \pi_{\omega', i}(\theta) u_{\omega', a_j}}{\sum_{\omega}\pi_{\omega, i}(\theta)} \geq 
    \frac{\sum_{\omega'} \theta_{\omega'} \pi_{\omega', i}(\theta) u_{\omega', a_i}}{\sum_{\omega}\pi_{\omega, i}(\theta)} - 2\frac{\varepsilon}{|\Omega|}  ~~~~\left(\text{as } |\hat{\pi}_{\omega, i} - \hat{\pi}_{\omega, j}| \leq \frac{\varepsilon}{|\Omega|}\right) \\
    \implies &\sum_{\omega'} \theta_{\omega'} \pi_{\omega', i}(\theta) u_{\omega', a_j} \geq 
    \sum_{\omega'} \theta_{\omega'} \pi_{\omega', i}(\theta) u_{\omega', a_i} - 2\frac{\varepsilon}{|\Omega|} \sum_{\omega'}\pi_{\omega', i}(\theta)\\
    \implies & \sum_{\omega'} \theta_{\omega'} (\pi_{\omega', i}(\theta) + \pi_{\omega', j}(\theta)) u_{\omega', a_j} \geq \sum_{\omega'} \theta_{\omega'} \pi_{\omega', i}(\theta) u_{\omega', a_i} \\
    &\qquad\qquad~~~~~~~~~~~~~~~~~~~~~~~~~~~~~~~~~~~~~~~~~~~~~~~~~~~~~~ + \sum_{\omega'} \theta_{\omega'} \pi_{\omega', j}(\theta) u_{\omega', a_j} - 2\frac{\varepsilon}{|\Omega|} \sum_{\omega'}\pi_{\omega', i}(\theta)  \\
    \implies & \max_{a \in A} \sum_{\omega'} \theta_{\omega'} (\pi_{\omega', i}(\theta) + \pi_{\omega', j}(\theta)) u_{\omega', a} \geq \sum_{\omega'} \theta_{\omega'} \pi_{\omega', i}(\theta) u_{\omega', a_i} \\&\qquad\qquad~~~~~~~~~~~~~~~~~~~~~~~~~~~~~~~~~~~~~~~~~~~~~~~~~~~~~~ + \sum_{\omega'} \theta_{\omega'} \pi_{\omega', j}(\theta) u_{\omega', a_j} - 2\frac{\varepsilon}{|\Omega|} \sum_{\omega'}\pi_{\omega', i}(\theta)
\end{align*}
Hence, we see that when we merge two signals that are in the same set, the value of the buyer for the experiment drops by at most $2\varepsilon \sum_{\omega'}\pi_{\omega', i}(\theta)$, where $\pi_{\cdot, i}(\theta)$ is the signal with the smallest sum. Let $\pi_{\cdot, i+j}$ be the merged signal and assume that for $\omega$ we have $\hat{\pi}_{\omega,i} \leq \hat{\pi}_{\omega,j}$. Then, it holds that $\hat{\pi}_{\omega,i} \leq \hat{\pi}_{\omega, i+j} \leq \hat{\pi}_{\omega,j}$ because for any four positive numbers $a, b, c, d$ with $\frac{a}{b} \leq \frac{c}{d}$ we have that $\frac{a}{b} \leq \frac{a+c}{b+d} \leq \frac{c}{d}$. Thus, we see that the merged signal will remain in the same set of the partition that the two original ones were. Let $\pi_{\cdot, 1}, \ldots, \pi_{\cdot, N}$ be the signals that are in some set $P$. Our previous discussion shows that we can merge $\pi_{\cdot, 1}, \pi_{\cdot, 2}$ to create $\pi_{\cdot, 1+2}$, then merge $\pi_{\cdot, 1+2}$ with $\pi_{\cdot, 3}$ and so on. Importantly, all these signals will remain in $P$ and the amount by which they decrease the value of the experiment is at most $2\frac{\varepsilon}{|\Omega|}\sum_{i \in P}\sum_{\omega \in \Omega}\pi_{\omega, i}(\theta)$. If we do that for all sets $P$, the total decrease in the value is at most $2\frac{\varepsilon}{|\Omega|}\sum_{i}\sum_{\omega \in \Omega}\pi_{\omega, i}(\theta) = 2\varepsilon$
\end{prevproof}

We now argue that when we merge two signals, no type can value any experiment more than he did before.

\begin{claim}\label{clm:merge drops value}
 Let $E({\theta})$ be the initial experiment that is offered to type $\theta$ and $E'({\theta})$ the experiment that is offered to $\theta$ after the merge. Then, for any type $\theta' \in \Theta$ it holds that $V_{\theta}(E'({\theta}')) \leq V_{\theta}(E({\theta}'))$.
 \end{claim}

 \begin{prevproof}{Claim}{clm:merge drops value}
Consider the first time that two signals $s_i, s_j$ of $E({\theta}')$ are merged. Then, the value of $\theta$ for this new experiment is
\begin{align*}
V_{\theta}(E'({\theta}')) &=  \sum_{l \neq i,j} \max_a \sum_{\omega} \theta_{\omega} \pi_{\omega, l}(E(\theta')) u_{\omega, a} +  \max_a \sum_{\omega} \theta_{\omega} (\pi_{\omega, i}(E(\theta')) 
+ \pi_{\omega, j}(E(\theta'))) u_{\omega, a} \\
&\leq \sum_{l \neq i,j} \max_a \sum_{\omega} \theta_{\omega} \pi_{\omega, l}(E(\theta')) u_{\omega, a} \\&
\qquad\qquad~~~~~~~~~~~~~~~~~~~+  \max_a \sum_{\omega} \theta_{\omega} \pi_{\omega, i}(E(\theta'))  u_{\omega, a}  + \max_a \sum_{\omega} \theta_{\omega} \pi_{\omega, j}(E(\theta'))  u_{\omega, a} \\
&= V_{\theta}(E({\theta}'))
\end{align*}
Continuing inductively, we prove the claim.
\end{prevproof}

So far we have established the existence of a menu $\MM'$ whose number of signals is significantly smaller than the initial one. However, if we do not have access to $\MM$ we cannot compute $\MM'$. Lemma~\ref{lem:signals rounding} shows that we can overcome this issue. By ``rounding'' the entries of the experiment so that for any type the value this experiment generates does not change much. Now, since there is a  small number of signals, and the size of every experiment depends only on $\varepsilon$ and $|\Omega|$, we can do an exhaustive search over the discretized entries.
\begin{lemma}\label{lem:signals rounding}
Let $\{E(\theta)\}_{\theta \in \Theta}$ be a set of experiments, where each $E(\theta)$ uses signals from $S$. We also let $0 < \delta < 1$ be a given number such that $1/\delta \in \N$. Then, we can create a set of experiments $\{E'(\theta)\}_{\theta \in \Theta}$ that uses signals from $S'$ with $|S'| = |S|$, such that $E'(\theta)$ is a valid experiment with $\pi_{\omega, s'}(E'(\theta))$ being a multiple of $\delta$ %= c_{\omega, s'} \cdot \delta$, where $c_{\cdot, \cdot}$ is an integer
for all $\omega \in \Omega, s' \in S', \theta \in \Theta$, and  $|V_{\theta}(E'(\theta')) - V_{\theta}(E(\theta')) | \leq \delta |S|$,  for all $\theta, \theta' \in \Theta$.
\end{lemma}

\begin{prevproof}{Lemma}{lem:signals rounding}
Fix some $\theta \in \Theta$ and consider any $\theta' \in \Theta$. We create $E'(\theta')$ by rounding every entry of $\pi_{\cdot, \cdot}(E(\theta'))$ to multiples of $\delta$ in such a way that $\sum_{i'} \pi_{\omega, i'} = 1, \forall \omega \in \Omega$. Consider a particular signal $s_k$ of $E(\theta')$ that results in $\theta$ taking action $a(s_k)$. After the rounding step, the rounded signal $s'_k$ might lead $\theta$ to take a different action $a(s'_k)$. Thus, we have
\begin{align*}
    V_\theta(E'({\theta}'))& = \sum_{s_k' \in S'}\sum_{\omega \in \Omega} \theta_\omega \pi_{\omega, k'}(E'(\theta')) u_{\omega, a(s_k')} \geq 
    \sum_{s_k' \in S'}\sum_{\omega \in \Omega}\theta_\omega \pi_{\omega, k'}(E'(\theta')) u_{\omega, a(s_k)} \\
   & \geq \sum_{s_k \in S} \sum_{\omega \in \Omega}\theta_\omega (\pi_{\omega, k}(E(\theta')) -\delta)u_{\omega, a(s_k)} \geq 
    V_\theta(E(\theta')) - \delta |S|
\end{align*}
The other direction is proved similarly.
\end{prevproof}

One construction that will be useful in our proofs is the $\varepsilon$-IC to IC transformation. Lemma~\ref{lem: eps-bic to bic} shows that if we have a menu whose IC, IR constraints are violated by at most $\varepsilon$, we can modify the prices so that it becomes IC, IR and has negligible {$O(\sqrt{\varepsilon})$} revenue loss. {The construction is based on a technique developed~\cite{DaskalakisW12,cai2021efficient} and frequently used in the Mechanism Design literature. In the single agent setting, the idea is to offer a small multiplicative discount to all types to make sure that if they want to deviate to some other experiment, this will not be much cheaper than the one they were buying in the initial $\varepsilon$-IC menu.}

\begin{lemma}\label{lem: eps-bic to bic}
    Let $\MM = \{E_i, t(E_i)\}_{i \in [k]}$ be a menu with $k$ experiments. Suppose that the IC, IR constraints are violated by at most $\varepsilon$. Then, we can compute a new set of prices $\{\tilde{t}(E_i)\}_{i \in [k]}$ such that the menu $\widetilde{\MM} = \{E_i, \tilde{t}(E_i)\}_{i \in [k]}$ is IR, IC and $\rev(\widetilde{\MM}) \geq (1-\sqrt{\varepsilon})\rev(\MM) - \sqrt{\varepsilon} - \varepsilon$, in time $O(k)$. 
\end{lemma}

\begin{prevproof}{Lemma}{lem: eps-bic to bic}
    Let $\tilde{t}(E_i) = (1-\eta) t(E_i) - \varepsilon$, where $\eta > 0$. We immediately see that all the IR constraints are now satisfied. Consider a type $\theta$ who buys experiment $E$ under the original prices. Since the IC constraints are violated by at most $\varepsilon$, we know that
$$ V_{\theta}(E) - t(E) \geq  V_{\theta}(E') - t(E') - \varepsilon,  ~ \forall E' \in \MM.$$
    Now suppose that $\theta$ prefers $E''$ under the new prices. Then
$$V_{\theta}(E'') - (1-\eta) t(E'') \geq  V_{\theta}(E) - (1-\eta)t(E).$$
    Choosing $E'$ to be $E''$ in the first inequality and  combining the two inequalities, we have that
    \begin{align*}
        V_{\theta}(E'') - (1-\eta) t(E'') \geq V_{\theta}(E'') - t(E'') - \varepsilon +\eta t(E) \implies        t(E) - t(E'') \leq \frac{\varepsilon}{\eta}
    \end{align*}
    Hence, for the revenue we have
    $\rev(\widetilde{\MM}) \geq (1-\eta)\rev(\MM) - \varepsilon - \frac{\varepsilon}{\eta}$. By picking $\eta = \sqrt{\varepsilon}$ we get the result.
\end{prevproof}

We are now ready to present a sketch of the proof for Theorem~\ref{thm:num signals approx}. Assume that we start with the optimal menu $\MM^*$. By Lemma~\ref{lem:merge signals}, Claim~\ref{clm:merge drops value}, and Lemma~\ref{lem:signals rounding}, we know that we can modify the experiments in $\MM^*$ so that they use only discretized signals. %whose number depends only on $|\Omega|, \varepsilon$.
Moreover, the new menu is approximately IC and IR. We then apply Lemma~\ref{lem: eps-bic to bic} to obtain a menu that is IC and IR by sacrificing  a negligible amount of revenue. Finally, to compute the collection of action sets that the types will choose after receiving the signals, we query BR oracle on all the possible discretized distributions where the signals are drawn from.\\

\begin{prevproof}{Theorem}{thm:num signals approx}
Let $\{E(\theta)\}_{\theta \in \Theta}$ be a set of experiments that use signals from $S$ and $\{t(\theta)\}_{\theta \in \Theta}$ be the corresponding set of prices that form a valid IC, IR menu. Let $\{E'(\theta)\}_{\theta \in \Theta}$ be the set of experiments that is induced by first merging and then rounding the experiments of $\{E(\theta)\}_{\theta \in \Theta}$. By Lemma~\ref{lem:merge signals} and Lemma~\ref{lem:signals rounding}, we have that $V_{\theta}(E'({\theta})) \geq V_{\theta}(E({\theta})) - 2\varepsilon - \delta |S'|$, where $|S'| = \Theta\left( (|\Omega|/\varepsilon)^{|\Omega|})\right)$. We pick $\delta$ so that $\delta |S'| \leq \varepsilon$. Thus, $V_{\theta}(E'({\theta})) \geq V_{\theta}(E({\theta})) - 3\varepsilon$. Moreover, Claim~\ref{clm:merge drops value} and Lemma~\ref{lem:signals rounding} guarantee that $V_{\theta}(E'(\theta')) \leq V_{\theta}(E(\theta')) + \delta|S'| \leq V_{\theta}(E(\theta')) + \varepsilon$ for any other experiment $E'(\theta')$ . Hence, the IC and IR constraints for the menu $\{(E'(\theta), t(\theta))\}_{\theta \in \Theta}$ are violated by at most $4\varepsilon$.
Lemma~\ref{lem: eps-bic to bic} shows how we can transform the prices $\{t(\theta)\}_{\theta \in \Theta}$ to $\{t'(\theta)\}_{\theta \in \Theta}$ so that the revenue drops by at most $\Theta(\sqrt{\varepsilon})$, without modifying the experiments that are offered. Considering $\{E(\theta), t(\theta)\}_{\theta \in \Theta}$ to be the optimal menu concludes gives us the revenue guarantee. Consider some type $\theta \in \Theta$. Note that if we let $\delta = \varepsilon/|S'|$, since every column of the experiment consists of $|\Omega|$ entries, we have that there are at most $(1/\delta)^{|\Omega|} = O\left( \frac{|\Omega|^{|\Omega|^2}}{\varepsilon^{|\Omega|^2 + |\Omega|}}\right)$ different possible columns that we can send to $\theta$. Every such column induces a posterior distribution for $\theta$, so by querying the BR oracle $O\left( \frac{|\Omega|^{|\Omega|^2}}{\varepsilon^{|\Omega|^2 + |\Omega|}}\right)$ many times, we find the set of actions $A_{\theta}$ that this type will ever consider after receiving any of the signals from $S'$. %Note that since there are $|\Theta|$ many types, the number of queries to BR oracle is $O\left(|\Theta| \frac{|\Omega|^{|\Omega|^2}}{\varepsilon^{|\Omega|^2 + |\Omega|}}\right)$
\end{prevproof}

So far we have only shown a structural result about the existence of a menu that uses $O\left( \frac{|\Omega|^{|\Omega|^2}}{\varepsilon^{|\Omega|^2 + |\Omega|}}\right)$ signals and gives an additive $O(\sqrt{ \varepsilon})$-approximation to the optimal revenue. We now argue that we can use a LP (Figure~\ref{fig:alg implicit}) to find such a menu. This is done by combining the result of Theorem~\ref{thm:num signals approx} and modifying the LP we used in Section~\ref{sec:explicit model}. 

\begin{theorem}\label{thm:lp gives approx}
 For any $\varepsilon > 0$, any set $\Theta$ of types, given access to a BR oracle, we can use the LP in Figure~\ref{fig:alg implicit} to compute a menu $\MM$ that achieves $\rev(\MM) \geq \opt - O(\sqrt{\varepsilon})$. The number of queries to the BR oracle is $\poly\left(|\Theta|, \frac{|\Omega|^{|\Omega|^2}}{\varepsilon^{|\Omega|^2 + |\Omega|}}\right)$ and the running time is $\poly\left(|\Theta| , \frac{|\Omega|^{|\Omega|^2}}{\varepsilon^{|\Omega|^2 + |\Omega|}}\right)$. %the number of different signals of $\MM$ is $O\left(|\Theta| \frac{|\Omega|^{|\Omega|^2}}{\varepsilon^{|\Omega|^2 + |\Omega|}}\right)$. 
 Moreover, each experiment contains at most $O\left( \frac{|\Omega|^{|\Omega|^2}}{\varepsilon^{|\Omega|^2 + |\Omega|}}\right)$ many signals, and each type $\theta$ only chooses actions from a set of actions $A_\theta$ with size $O\left( \frac{|\Omega|^{|\Omega|^2}}{\varepsilon^{|\Omega|^2 + |\Omega|}}\right)$.
\end{theorem}

\begin{prevproof}{Theorem}{thm:lp gives approx} 
By Theorem~\ref{thm:num signals approx}, we know that we can construct a collection of actions sets $\{A_{\theta}\}_{\theta \in \Theta}$ in time $O\left(|\Theta| \frac{|\Omega|^{|\Omega|^2}}{\varepsilon^{|\Omega|^2 + |\Omega|}}\right)$ so that there exists a valid IC, IR menu $\MM'$ that only recommends actions from $A_{\theta}$ to any type $\theta$. The running time of this construction and the number of queries to the BR oracle are both $O\left(|\Theta| \frac{|\Omega|^{|\Omega|^2}}{\varepsilon^{|\Omega|^2 + |\Omega|}}\right)$. We assume set $A_\theta=\{a_{\tau_{\theta}(i)}\}_{i\in[|A_{\theta}|]}$ for every type $\theta$. %Let $\tau_{\theta}(i)$ be the action that $\theta$ takes when he receives signal $s_i$ from $E(\theta)$. 
Now consider the LP in Figure~\ref{fig:alg implicit}, which is a modified version of the one in Section~\ref{sec:explicit model}.

First we query BR oracle for every $\theta \in \Theta$ to figure out what $u(\theta)$ is for every type $\theta$. Since $\MM'$ is IC and IR, we know that this menu is a feasible solution for the LP ~\footnote{Some actions may not be recommended in the experiment for $\theta$ in $\M'$. Simply set $\pi_{\omega,i}(\theta)$ to $0$ for those actions.}.
However, in this case the second set of constraints may contain an arbitrarily large number of inequalities. Nevertheless, we can construct a polynomial time Separation Oracle that checks these constraints using the BR oracle, hence we can solve the LP with the Ellipsoid Algorithm. The Separation Oracle works as follows. Fix the value of the variables in the LP and consider two types $\theta, \theta'$ and some signal $s_i \in [|A_{\theta'}]|$. For the posterior of $\theta$ that this signal induces, we can query the BR oracle to figure out what the best action $a_j \in A$ is. Hence, we can find the ``tightest'' constraint regarding the variable $z_i(\theta, \theta')$ with one query. Since there are only  $O\left(|\Theta|^2 \frac{|\Omega|^{|\Omega|^2}}{\varepsilon^{|\Omega|^2 + |\Omega|}}\right)$  such variables, we need that many queries to the BR oracle to check if one of them is violated. For the other set of constraints, it is easy to check whether they are violated. So we can solve this LP in time $\poly\left(|\Theta|,  \frac{|\Omega|^{|\Omega|^2}}{\varepsilon^{|\Omega|^2 + |\Omega|}}\right)$. Observe that, as in Theorem~\ref{thm:alg explicit}, every feasible point of this LP is a responsive menu. Moreover, since there exists a feasible point that generates revenue at least $\opt - O(\sqrt{\varepsilon})$ we know that every optimal solution of the LP generates at least this much revenue.
\end{prevproof}

%Since $\theta'$ gets at most $O\left(\frac{|\Omega|^{|\Omega|^2}}{\varepsilon^{|\Omega|^2 + |\Omega|}}\right)$ signals, we can query BR oracle that many times to figure out the value of $\theta$ for the experiment of $\theta'$ and check whether this constraint is violated in $O\left( \frac{|\Omega|^{|\Omega|^2}}{\varepsilon^{|\Omega|^2 + |\Omega|}}\right)$ time. There are $|\Theta^2|$ many of them, so we can indeed construct a Separation Oracle for these constraints in time polynomial in  $|\Theta|$ and $\left( \frac{|\Omega|^{|\Omega|^2}}{\varepsilon^{|\Omega|^2 + |\Omega|}}\right)$.

\begin{figure}[h!]
\colorbox{MyGray}{
\begin{minipage}{1.1\textwidth} {
\noindent\textbf{Variables:}
\begin{itemize}
\item $\{\pi_{\omega,i}(\theta)\}_{\omega\in \Omega, i\in [|A_\theta|], \theta\in \Theta}$, denoting the experiments in the menu.
\item $\{t(\theta)\}_{\theta\in \Theta}$, denoting the prices of the experiments. 
\item $\{z_i(\theta, \theta')\}_{i \in [|A_{\theta'}|]\theta, \theta' \in \Theta}$, helper variables. $z_i(\theta, \theta')$ represents an upper bound of the conditional expected utility of signal $s_i$ from experiment $E(\theta')$ for type $\theta$.
\end{itemize}
\textbf{Linear Program:}
\begin{equation*}
\begin{array}{ll@{}ll}
\text{max}  & \displaystyle\sum\limits_{\theta\in \Theta} F(\theta)t(\theta)&\\
% \text{subject to} &\displaystyle\sum\limits_{\omega}\theta_{\omega} \pi_{\omega, i}(\theta)u_{\omega, a_{\tau_\theta(i)}} \geq 
%                     \displaystyle\sum\limits_{\omega}\theta_{\omega} \pi_{\omega, i}(\theta)u_{\omega, a_j},  & \forall \theta, \forall i \in [|A_{\theta}|], \forall a_j \in A ~~~(\text{OB})\\
     \text{s.t.}           &  \displaystyle\sum\limits_{i\in [|A_\theta|]}\displaystyle\sum\limits_{\omega\in \Omega} \theta_{\omega} \pi_{\omega, i}(\theta)u_{\omega, a_{\tau_{\theta}(i)}} - t(\theta) \geq
                    \displaystyle\sum\limits_{i\in [|A_{\theta'}|]}z_i(\theta, \theta') - t(\theta'),~~~ 
                &   \forall \theta, \theta'\in \Theta  ~~~(\text{IC})\\
                & z_i(\theta, \theta') \geq \displaystyle\sum_\omega \theta_\omega \pi_{\omega, i}(\theta') u_{\omega, a_j}, &\forall \theta, \theta' \in \Theta, \forall  i \in [|A_{\theta'}|], \forall a_j \in A\\
                & \displaystyle\sum\limits_{i\in [|A_\theta|]}\displaystyle\sum\limits_{\omega\in \Omega} \theta_{\omega} \pi_{\omega, i}(\theta)u_{\omega, a_{\tau_{\theta}(i)}} - t(\theta) \geq u(\theta),
                & \forall \theta\in \Theta ~~~(\text{IR})\\  
                & \displaystyle\sum\limits_{i\in [|A_\theta|]} \pi_{\omega, i} (\theta) = 1, & \forall \theta\in \Theta, \omega\in \Omega \\
                & \pi_{\omega, i}(\theta) \geq 0, & \forall \theta\in \Theta, \forall \omega\in \Omega, \forall i\in [|A_\theta|]
\end{array}
\end{equation*}}
\end{minipage}} \caption{A linear program to find an approximately revenue-optimal menu in the implicit model.}\label{fig:alg implicit}
\end{figure}

So far, the number of experiments in our constructions depends on the number of types $|\Theta|$. We show that the number of experiments needed  for an up-to-$\varepsilon$ optimal menu is independent of $|\Theta|$. %We first consider an arbitrary menu and prove that if we offer some type $\theta$ the same experiment that is offered to $\theta'$, where the total variation distance $d_{TV}(\theta, \theta')$ between $\theta$ and $\theta'$  is less than $\varepsilon$, then $V_\theta(E(\theta'))$ is close to $V_\theta(E(\theta))$.
{We achieve this by dropping experiments that are offered to types who are close in TV-distance. We show that this leads to a menu that preserves the revenue, is $O(|\Omega|\varepsilon)$-IC, IR and has $O\left(\frac{|\Omega|^{2|\Omega|}}{\varepsilon^{|\Omega|}} \right)$ different experiments. Finally, we apply the $\varepsilon$-IC to IC transformation to the modified menu.}

\begin{lemma}\label{lem:change offers theta}
Let $\mathcal{M} = \{E(\theta), t(\theta)\}_{\theta \in \Theta}$ be a menu of experiments. Then, for any $\theta, \theta' \in \Theta$ with $d_{TV}(\theta, \theta') \leq \varepsilon$ it holds that $V_{\theta}(E({\theta'})) - t(\theta') \geq V_{\theta}(E({\theta})) - t(\theta) - 2|\Omega|\varepsilon$.
\end{lemma}

\begin{prevproof}{Lemma}{lem:change offers theta}
Let $a(s_k)$ be the action that type $\theta$ takes upon receiving signal $s_k \in S_{\theta'}$ from $E(\theta')$ and $a'(s_k)$ the action that $\theta'$ takes upon reception of the same signal. Then we have that
\begin{align*}
    & V_\theta(E({\theta'})) =  \sum_{s_k \in S_{\theta'}} \sum_{\omega \in \Omega} \theta_{\omega}\pi_{\omega, k}(E(\theta')) u_{\omega, a(s_k)} \geq \sum_{s_k \in S_{\theta'}} \sum_{\omega \in \Omega} \theta_{\omega}\pi_{\omega, k}(E(\theta')) u_{\omega, a'(s_k)} \\ &~~~~~~~~~~~~~~~~~\geq \sum_{s_k \in S_{\theta'}}\left( \sum_{\omega \in \Omega} \theta'_{\omega}\pi_{\omega, k}(E(\theta')) u_{\omega, a'(s_k)} - \varepsilon\cdot \sum_{\omega\in \Omega} \pi_{\omega, k}(E(\theta')\right)\\
    \implies &   V_\theta(E({\theta'})) \geq V_{\theta'}(E({\theta'})) - |\Omega|\varepsilon
\end{align*}

Similarly, we have that $ V_{\theta'}(E({\theta})) \geq V_{\theta}(E({\theta})) - |\Omega|\varepsilon$. 

Since $\mathcal{M}$ is IC, it holds that $V_{\theta'}(E({\theta'})) - t(\theta') \geq V_{\theta'}(E({\theta})) - t(\theta)$. Thus, using the inequalities above we have that
$$
    V_{\theta'}(E({\theta'})) - t(\theta') \geq V_{\theta'}(E({\theta})) - t(\theta) \geq V_{\theta}(E_{\theta}) - t(\theta) - |\Omega|\varepsilon,$$
    and
$$ V_{\theta}(E({\theta'})) - t(\theta')  \geq V_{\theta'}(E({\theta'})) - t(\theta') - |\Omega|\varepsilon, $$ which implies
$$    V_{\theta}(E({\theta'})) - t(\theta') \geq
    V_{\theta}(E({\theta})) - t(\theta) - 2|\Omega|\varepsilon.$$
    \end{prevproof}

We are now ready to prove that we can create a menu that offers a small number of experiments and loses negligible revenue compared to $\opt$. We do that in two steps, since we are dealing with an action space and type space that are arbitrary. The first step is to shrink the action space that we are considering. In order to do that, we use Theorem~\ref{thm:num signals approx} that guarantees the existence of a menu which loses negligible revenue and only considers actions from smaller action spaces. The next step is to divide the state space into regions in which all the types are within $\varepsilon$ in TV-distance. Lemma~\ref{lem:change offers theta} shows us that if we consider offering a single experiment to all the types in the same region, their values for the new experiment will not change much compared to the one they were getting. Finally, we apply Lemma~\ref{lem: eps-bic to bic} to solve the issue that the menu resulting from dropping experiments might not be IC, IR.

\begin{theorem}\label{thm:arbitrary theta}
    Consider an environment with a type space $\Theta$, action space $A$ and state space $\Omega$.
    Then, given some $\varepsilon >0$ and access to a BR oracle we can find a menu $\mathcal{M}$ that generates revenue at least $OPT - O(\sqrt{\varepsilon})$ and offers at most $O\left(\frac{|\Omega|^{2|\Omega|}}{\varepsilon^{|\Omega|}} \right)$ experiments, in time $\poly\left(|\Theta|, \frac{|\Omega|^{|\Omega|^2}}{\varepsilon^{|\Omega|^2 + |\Omega|}}\right)$.
\end{theorem}

\begin{prevproof}{Theorem}{thm:arbitrary theta}
    Let $\varepsilon_1, \varepsilon_2 > 0$ be some constants that will be specified later. Also, let $\MM^*$ be the optimal menu for this environment. We first observe that Theorem~\ref{thm:lp gives approx} shows that we can get a menu $\MM$ that sends at most $O\left( \frac{|\Omega|^{|\Omega|^2}}{\varepsilon_1^{|\Omega|^2 + |\Omega|}}\right)$ different signals and generates revenue $\rev{(\MM)} \geq \opt - O(\sqrt{\varepsilon_1})$ in time $\poly\left(|\Theta|,  \frac{|\Omega|^{|\Omega|^2}}{\varepsilon_1^{|\Omega|^2 + |\Omega|}}\right)$. The number of experiments that $\MM$ offers is at most $|\Theta|$. In order to get rid of this dependence on $|\Theta|$ we partition the types into regions in which all the types are within $\varepsilon_2$ in TV-distance. One way to do that is to round every $\theta_\omega$ to multiples of $\varepsilon_2/|\Omega|$, so after this step there will be at most $O\left((|\Omega|/\varepsilon_2)^{|\Omega|}\right)$ different types. Note that this guarantees that all the types that are rounded to the same type are within $\varepsilon_2$ TV-distance. Let $\tilde{\Theta}$ be the set of rounded types. We consider a new menu $\MM'$ that offers all the types $\theta \in \Theta$ who are rounded to the same $\tilde{\theta} \in \tilde{\Theta}$ the most expensive expensive experiment among $E(\theta)$ at price $t(\theta)$. It is clear that if the types are willing to buy the experiments that $\MM'$ offers them then $\rev{(\MM')} \geq \rev{(\MM)}$. However, it could be the case that they actually deviate and buy some other experiment in the menu. Consider some type $\theta$ who was getting $E(\theta)$ from $\MM$ and is now getting $E(\theta')$ from $\MM'$. Lemma~\ref{lem:change offers theta} shows that $V_{\theta}(E({\theta'})) - t(\theta') \geq V_{\theta}(E({\theta})) - t(\theta) - 2|\Omega|\varepsilon_2$. Moreover, for any other experiment $E'' \in \MM'$ we know that $V_\theta(E'') - t(E'') \leq V_\theta(E(\theta)) - t(E(\theta)) \leq V_\theta(E(\theta')) - t(E(\theta')) - 2|\Omega|\varepsilon_2$. Hence, we see that the IC, IR constraints for $\MM'$ are violated by at most $2|\Omega|\varepsilon_2$. Now Lemma~\ref{lem: eps-bic to bic} shows how to modify the prices of $\MM'$ to create a menu that is exactly IC, IR and loses at most $O(\sqrt{|\Omega|\varepsilon_2})$ revenue compared to $\MM'$. We call this new menu $\widetilde{\MM}$. Plugging the values in, we see that $\rev{(\widetilde{\MM)}} \geq \opt - O\left(\sqrt{|\Omega|\varepsilon_2}\right) - O(\sqrt{\varepsilon_1})$. Thus, we see that we can set $\varepsilon_1 = \varepsilon, \varepsilon_2 = \frac{\varepsilon}{|\Omega|}$ and this would guarantee revenue at least $\opt-O(\sqrt{\varepsilon})$. The number of signals is $O\left( \frac{|\Omega|^{|\Omega|^2}}{\varepsilon^{|\Omega|^2 + |\Omega|}}\right)$ and there are $O\left(\frac{|\Omega|^{2|\Omega|}}{\varepsilon^{|\Omega|}} \right)$ different experiments.
\end{prevproof}

\subsection{Implicit Model with Succinct Description}\label{sec:hardness without BRO}
In this section, we consider a setting where the model has a succinct implicit description. We show that no algorithm  can obtain even a constant factor approximation to the optimal revenue for this setting, unless P = NP. To be more precise, we consider the following problem.\\\\  \textbf{ Information Pricing SAT (IP-SAT):} find the revenue-optimal menu in the following setting:
\begin{itemize}
    \item State space $\Omega$, type space $\Theta$. 
    \item for each state $\omega$, $\Phi_{\omega}$ is a boolean formula in CNF over variables in $X = \{x_1, \ldots, x_n\}$
    \item Action space $A$: all the possible truth assignments of the variables in $X$
    \item $u_{\omega, a} = \frac{\text{$\#$ satisfied clauses of $\Phi_\omega$ with assignment $a$}}{\text{$\#$ clauses of $\Phi_\omega$}}$
\end{itemize}

 IP-SAT is a hard problem for the buyer in general, as maximizing his net utility requires solving an NP-hard problem. We show here that designing an approximately revenue-optimal menu for IP-SAT is also computationally intractable for the seller. Of course, it is not even clear what the optimal menu looks like in general for IP-SAT as we only have a limited characterization of the optimal menu. In Theorem~\ref{thm:sat reduction}, we construct a special family of IP-SAT instances with $2$ states and $1$ buyer type, and show how to reduce SAT to it. %Clearly, despite the fact that ED-SAT has a succinct representation, the action space of the agent is exponentially big in this representation. What makes this problem interesting is that $\Phi_{\omega_1}, \Phi_{\omega_2}$ are designed in a way that knowing the state of the world actually allows the buyer to find an optimal assignment. It is not very hard to see that since the agent has only one type, the seller should send him the fully informative experiment to maximize her revenue, i.e. just reveal to him the state of the world. However, as Theorem~\ref{thm:sat reduction} shows the task of figuring out the optimal price for this experiment is NP-hard.

\begin{theorem}\label{thm:sat reduction}
For any constant $\varepsilon>0$, there does not exit a polynomial time algorithm $\mathcal{A}$ that computes an menu with revenue at least $(1/2+\varepsilon)\opt-\frac{\varepsilon}{2m+4}$ in the IP-SAT problem with $m+2$ clauses,  unless $P = NP$. 
\end{theorem}
\begin{prevproof}{Theorem}{thm:sat reduction}
        Let $\Phi = C_1 \land  \ldots \land C_m$ be any SAT instance over variables $x_1, \ldots, x_n$. We show that given $\mathcal{A}$ we can decide whether $\Phi$ is satisfiable. We create the following IP-SAT instance: there are two states $\omega_1, \omega_2$, a single type $\theta=(\frac{1}{2},\frac{1}{2})$ and we set $\Phi_{\omega_1} = (C_1 \lor y) \land \ldots \land (C_m \lor y)\land (x_1 \lor y) \land (\lnot x_1 \lor y), \Phi_{\omega_2} = (C_1 \lor \lnot y)\land \ldots \land (C_m \lor \lnot y)\land (x_1 \lor \lnot y) \land (\lnot x_1 \lor \lnot y)$ to be the two SAT instances with $m+2$ clauses that the buyer faces. At each state, the actions are the possible assignments of the variables $x_1, \ldots, x_n, y$. Let $\Psi$ be some boolean formula in CNF and $a$ some assignment of its variables. We define $z_a(\Psi)$ to be the number of clauses in $\Psi$ that $a$ satisfies and $w(\Psi)$ the total number of clauses in $\Psi$. Then, at each state $\omega$ when the agent chooses assignment $a$ we define his ex-post utility to be $u_{\omega, a} = \frac{z_a(\Phi_\omega)}{w(\Phi_\omega)}$.

       From Bergemann et al.~\cite{bergemann2018design}, we know the optimal menu should contain only the fully informative experiment $E^*$, and the price for this experiment is $V_{\theta}(E^*) - u(\theta)$.~\footnote{The fully informative experiment simply sends out a signal to reveal the state.} Clearly, $V_{\theta}(E^*) = 1$, because the buyer can pick $y$ to be $T$ and $F$ in states $\omega_1$, $\omega_2$ respectively and satisfy all clauses. We now focus our attention on $u(\theta)$. Assume that without receiving any information the buyer decides to set $y=T$. This is w.l.o.g. since it is symmetric with the case he decides to set $y=F$. When the state is $\omega_1$, he satisfies all the clauses. According to his prior, this happens $1/2$ of the time, so we see that so far $u(\theta) \geq 1/2$. Let us consider which assignment he should pick when the state is $\omega_2$. Observe that no matter which value he picks for $x_1$, he will always satisfy exactly one of the last two clauses in $\Phi_2$. Hence, for variables $x_1, \ldots, x_n$ he better pick the assignment $a$ that maximizes $z_a(\Phi)$. Let $k = \max_a z_a(\Phi)$. Then $u(\theta) = 1/2 + (k+1)/(2m+4 )= \frac{m+k+3}{2m+4}$. Hence, the optimal revenue is $V_\theta(E^*) - u(\theta) = 1 -  \frac{m+k+3}{2m+4} = \frac{m-k+1}{2m+4}$. If $\Phi$ is satisfiable, then $k = m$, so $\opt = \frac{1}{2m+4}$. If $\Phi$ is not satisfiable then $k \leq m-1$ so $\opt \geq \frac{2}{2m+4}$. Now assume that there is such an algorithm $\mathcal{A}$ and denote $\rev(\mathcal{A})$ the revenue generated by the mechanism output by $\mathcal{A}$. If $\Phi$ is not satisfiable, since $\opt \geq \frac{2}{2m+4}$, it must be that $\rev(\A) > \frac{1}{2m+4}$. On the other hand, if $\Phi$ is not satisfiable we have that $\rev(\A) \leq \frac{1}{2m+4}$. Hence, the existence of $\A$ allows us to distinguish between satisfiable and unsatisfiable SAT formulas.
\end{prevproof}

\begin{remark}
Since $m$ is linear in the description length of the problem, given access to a BR oracle, for any $\varepsilon > 0$ we can set $\varepsilon' = c(m\varepsilon)^2$, for some appropriate constant $c > 0$, and then apply Theorem~\ref{thm:lp gives approx}. Since $|\Omega| = 2$ and $|\Theta| = 1$, the revenue we get is at least $\opt - \varepsilon$ and the running time is $O\left(\frac{1}{(m\varepsilon)^{10}}\right)$. However, this does not contradict with the result of Theorem~\ref{thm:sat reduction}, since in this setting the BR oracle solves an NP-hard problem.
\end{remark}

\section{Multi-Agent Setting}\label{sec:multi-agent}
In this section, we consider a multi-agent generalization of the model by Bergemann et al. \cite{bergemann2018design}. 
More specifically, we assume that there are $n$ buyers who are interested in acquiring extra information from the seller and each buyer's ex-post utility only depends on the state of the world and his own action. We further assume that the types of the buyers are drawn independently from their own type distributions. If there is no competition among them, the solution to the problem follows immediately from the single-agent setting, since the seller can offer each agent his optimal menu separately. Thus, we focus on a more interesting case where the buyers are competitors and only \textit{one} of them can receive an informative signal. 

\vspace{-.1in}
\paragraph*{Input Model and New Notation} We first need to introduce some new notation. We use $\Theta^i$ to denote the type space of buyer $i$ and $F^i(\theta^i)$ to denote the probability that buyer $i$'s type is $\theta^i$. We use $\Theta$ to denote the set of all type profiles and $F(\theta)$ to denote $\times_{i\in[n]} F^i(\theta^i)$. We assume the action space $A$ is the same for each buyer $i$, but the ex-post utility $u^i_{\omega,a}$ for choosing action $a$ under state $\omega$ may be different for different buyers. We consider the \emph{explicit model}, that is, for each buyer $i$, both $F^i$ and the ex-post utility matrix $U^i=\{u^i_{\omega,a}\}_{\omega\in \Omega, a\in A}$ are given as input. We use $u^i(\theta^i)$ to denote the base utility of buyer $i$ for choosing the best action under distribution $\theta^i$. 

\vspace{-.1in}
\paragraph*{Interaction between the Seller and Buyers} The interaction happens in the following order:
\begin{enumerate}
    \item The seller commits to a mechanism $\left\{\left(\Pi(\theta)=\left(\Pi^1(\theta),\ldots,\Pi^n(\theta)\right)\right\},\left\{t(\theta)=\left(t^1(\theta),\ldots, t^n(\theta)\right)\right)\right\}_{\theta \in \Theta}$, and announces the mechanism to all buyers.
    \item The types of the buyers $\theta=(\theta^1,\ldots, \theta^n)$ are realized.
    \item Each buyer $i$ \textbf{privately} submits his type $\theta^i$ to the seller.
    \item The seller chooses buyer $i$ as the winner with probability $p^i(\theta)$.
    \item The seller observes the state of the world $\omega$ and sends buyer $i$ a signal $s$ according to the signal{ing} scheme $\Pi^i(\theta)$ and charges buyer $i$ price $t^i(\theta)$.
    \item Each buyer $i$ chooses an action $a^i$ and receives ex-post utility $u^i_{\omega, a^i}$.
\end{enumerate}

%he buyers submit privately a type profile $\theta=(\theta^1,\ldots, \theta^n)$ and the seller picks buyer $i$ as the winner with probability $p^i(\theta)$. After the state of the world $\omega$ is realized, she sends her a signal $s$ drawn from the distribution $\pi^i_{\omega, \cdot}(\theta)$. 
There are some subtle issues in our model that require further clarification. The most important of them being the following. After the winner has been chosen, does he  observe the signaling scheme $\Pi^i(\theta)$ that the seller uses to generate the signal $s$? In this work, we consider the setting where the signaling scheme $\Pi^i(\theta)$ is \textbf{not revealed} and the winner only observes the realized signal.~\footnote{One may worry that the winner can obtain extra information from the price $t^i(\theta)$. To avoid this, we will design a mechanism so that the price for buyer $i$ only depends on $i$'s type $\theta^i$. This is without loss of generality, as we can simply set the price to be $\E_{\theta^{-i}}[t^i(\theta)]$.} Some remarks are in order. Firstly, the seller may want to preserve the privacy of the buyers, and revealing $\Pi^i(\theta)$ allows the winner $i$ to infer the other buyers' priors. Secondly, hiding the implemented signaling scheme $\Pi^i(\theta)$ from the winner allows the seller to design a mechanism with less stringent IC constraints and thus generates higher revenue~for~the~seller. This is because the winner does not know the exact experiment he is getting, if he wants to deviate from the recommendation he must map the same signal to the same action for all potential experiments that he may win. Therefore, he would map the signal to an action that induces the highest \emph{expected utility}, where the expectation is over the other bidders' types and the chosen experiment. On the other hand, if the buyer knew which experiment the signal is drawn from, he could use a mapping that is the best for each particular experiment.

Our goal in this section is to design a polynomial time algorithm to find the Bayesian Incentive Compatible (BIC) and Interim Individually Rational (IIR) mechanism that achieves the highest revenue among all BIC and IIR mechanisms for our model. It is not hard to see that Lemma~\ref{lem:responsive menu} generalizes to our multi-agent setting. We begin by introducing an extension of the LP in Figure~\ref{fig:alg explicit} to the multi-agent setting. Define $\Pi^i(\theta) = \left( \pi^i_{\omega, j}(\theta) \right)_{\omega \in \Omega, j \in [m] }$, $p^i(\theta)$, and $t^i(\theta)$ as the decision variables for every buyer $i$ and type profile $\theta$. Recall that $m=|A|$. 

%We assume that the seller commits to a signaling scheme $\{\Pi(\theta)\}_{\theta \in \Theta}$ but when she picks the winner, she only sends her the signal without revealing the actual $\Pi(\theta)$ that was used. Hence, agent $i$ can only observe the expectation of the experiment that was used. This means that the Obedience constraint for every buyer need only be satisfied on average. We also assume that the agents are risk-neutral, i.e. the IR constraints are satisfied on average.\\

% Based on the way the reports of the agents are submitted to the seller, we consider two different models in the multi-agent setting:
% \begin{enumerate}
%     \item The types of the agents are submitted simultaneously and \textit{privately}.
    
%     \item The types of the agents are submitted simultaneously and \textit{publicly}.
% \end{enumerate}

% We emphasize that there is a subtle difference in these two settings. Consider some agent $i$. Essentially, when she sees the reports of the other agents she gets an extra signal that informs her about the experiment she will get from the seller. On the other hand, when she does not observe the other reports, she only knows a distribution over the experiments the seller will send. As a result, in the former case the obedience constraints have to be satisfied for every type profile, whereas in the latter case the obedience constraints only have to be satisfied on average. As it will be clear in the subsequent subsections, this plays an important role in the computational complexity of our approach to solve the problem.

\begin{equation*}
\begin{array}{ll@{}ll}
\text{max}  & \displaystyle\sum\limits_{\theta\in \Theta} F(\theta) \displaystyle\sum\limits_{i\in[n]} t^i(\theta)&\\
\text{s.t}        
                %\displaystyle\sum\limits_{\theta^{-i}}F^{-i}(\theta^{-i})\displaystyle\sum\limits_{\omega}\sum_{j}\theta^i_{\omega} \pi^i_{\omega, j}(\theta^i, \theta^{-i})u_{\omega, a_j} \geq 
                    %\displaystyle\sum\limits_{\-i} F(\theta^{-i})\displaystyle\sum\limits_{\omega}\theta^i_{\omega} \pi^i_{\omega, a}(\theta^i, \theta^{-i})u_{\omega, a'},  &\forall i, \forall \theta^i,  \forall a, \forall a'  ~~~(\text{Obedience})\\

        %  &\displaystyle\sum\limits_{\theta^{-i}}F^{-i}(\theta^{-i})  \displaystyle\sum\limits_{\omega\in \Omega} \theta^i_{\omega} \pi^i_{\omega, j}(\theta^i, \theta^{-i})u^i_{\omega, a_j} \geq
                    
        %         \displaystyle\sum\limits_{\theta^{-i}}F^{-i}(\theta^{-i}) \displaystyle\sum\limits_{\omega\in \Omega} \theta^i_{\omega} \pi^i_{\omega, j}(\theta^i, \theta^{-i})u^i_{\omega, a_k},
        %         &  ~~~ \forall i, \forall j, k,
        %         \forall \theta^i ~ (\text{OB})\\
                
    & \displaystyle\sum\limits_{\theta^{-i}}F^{-i}(\theta^{-i}) \left( \displaystyle\sum\limits_{\substack{\omega\in \Omega,\\j\in[m]}} \theta^i_{\omega} \pi^i_{\omega, j}(\theta^i, \theta^{-i})u^i_{\omega, a_j}+(1-p^i(\theta))u^i(\theta^i) - t^i(\theta^i, \theta^{-i}) \right)\geq \\
                    
                & \displaystyle\sum\limits_{\theta^{-i}}F^{-i}(\theta^{-i})\left(\displaystyle\sum\limits_{j\in[m]} z^i_j(\theta^i, \tilde{\theta^i}, \theta^{-i}) + (1-p^i(\tilde{\theta^{i}}, \theta^{-i}))u^i(\theta^i) - t^i(\tilde{\theta^{i}}, \theta^{-i})\right),
                ~~&   \forall i, 
                \forall \theta^i, \tilde{\theta}^i~ (\text{BIC})\\
                
                 & \displaystyle\sum\limits_{\theta^{-i}}F^{-i}(\theta^{-i}) z^i_j(\theta^i, \tilde{\theta^i}, \theta^{-i}) \geq  
                 \displaystyle\sum\limits_{\theta^{-i}}F^{-i}(\theta^{-i})\displaystyle\sum\limits_{\omega} \theta^i_\omega \pi^i_{\omega, j}(\tilde{\theta^i}, \theta^{-i})u_{\omega, a_k}, & \forall i, \forall j, k, \forall \theta^i, \tilde{\theta^i} \\
                
               &\displaystyle\sum\limits_{\theta^{-i}}F^{-i}(\theta^{-i}) \left( \displaystyle\sum\limits_{\substack{\omega\in \Omega,\\j\in[m]}} \theta^i_{\omega} \pi^i_{\omega, j}(\theta)u_{\omega, a_j}+(1-p^i(\theta))u^i(\theta^i) - t^i(\theta) \right)\geq u^i(\theta^i),                & \forall i, \theta^i ~~~(\text{IIR})\\

                & \displaystyle\sum\limits_{j\in[m]} \pi^i_{\omega, j} (\theta) = p^i(\theta),  \qquad\qquad \forall i, \forall \theta, \forall \omega ~~~(\text{feasibility})&\\
                & \displaystyle\sum\limits_{i\in[n]} p^i(\theta) \leq 1, ~~~\qquad\qquad\qquad \forall \theta ~~~(\text{feasibility})&\\
                & \pi^i_{\omega, j}(\theta) \geq 0,  ~~~~~~~\qquad\qquad\qquad \forall i, \forall \theta, \forall \omega, \forall j ~~~(\text{feasibility})&
\end{array}
\end{equation*}

Observe that the number of variables is exponential in $n$ and the number of constraints is exponential in both $n$ and $m$. There is no hope to solve this LP in polynomial time. The main challenge is how to remove the exponential dependence on $n$. To overcome this obstacle, we use a method that is powerful in the study of multi-item auctions, that is, rewriting the LP using a more succinct representation of the mechanism known as the reduced form~\cite{CaiDW12a, AlaeiFHHM12, CaiDW12b,CaiDW13a,CaiDW13b}. We first define the reduced form of a mechanism.

\begin{definition}[Reduced Form]~\label{def:reduced form}
%Given a mechanism $\M=\left(\left\{\Pi(\theta)\right\}_{\theta\in \Theta}, \{t(\theta)\}_{\theta\in \Theta}\right)$, we define its reduced form $\left(\left\{\hat{\Pi}^i(\theta^i)\right\}_{i\in[n],\theta^i\in \Theta^i}, \{\hat{t}^i(\theta^i)\}_{i\in[n],\theta^i\in \Theta^i}\right)$, where $\hat{\Pi}^i(\theta^i)=\{\hat{\pi}^i_{\omega,j}(\theta^i)\}_{\omega\in\Omega,j\in[m]}$ and $\hat{\pi}^i_{\omega,j}(\theta^i)=\E_{\theta^{-i}}[\pi^i_{\omega,j}(\theta)]$ for each state $\omega$ and $j\in[m]$, and $\hat{t}^i(\theta^i)=\E_{\theta^{-i}}[t^i(\theta)]$. We use $\PP(F)$ to denote the set of all reduced forms for a particular type distribution $F$.
Given a mechanism $\M=\left(\left\{\Pi(\theta)\right\}_{\theta\in \Theta}, \{t(\theta)\}_{\theta\in \Theta}\right)$, we define its reduced form $\left\{\hat{\Pi}^i(\theta^i)\right\}_{i\in[n],\theta^i\in \Theta^i}$, where $\hat{\Pi}^i(\theta^i)=\{\hat{\pi}^i_{\omega,j}(\theta^i)\}_{\omega\in\Omega,j\in[m]}$ and $\hat{\pi}^i_{\omega,j}(\theta^i)=\E_{\theta^{-i}}[\pi^i_{\omega,j}(\theta)]$ for each state $\omega$ and $j\in[m]$, and its interim prices $\{\hat{t}^i(\theta^i)\}_{i\in[n],\theta^i\in \Theta^i}$, where $\hat{t}^i(\theta^i)=\E_{\theta^{-i}}[t^i(\theta)]$. We use $\PP(F)$ to denote the set of all reduced forms for a particular type distribution $F$.
\end{definition}

It is not hard to see that $\PP(F)$ is a closed convex set, as the set of all mechanisms is clearly closed and convex, and $\PP(F)$ is simply a linear transformation of that set. {Intuitively, the reduced form is the ``expected experiment and price'' that the each buyer believes he will be allocated when his type is realized, and the expectation is taken over the randomness of the other buyers' types.}
\begin{lemma}
    For any type distribution $F$, $\PP(F)$ is a closed convex set. 
\end{lemma}

The LP in Figure~\ref{fig:alg multi-agent} searches for the reduced form of the revenue-optimal mechanism. Notice that the size of the reduced-form LP is substantially smaller than the original one, and the number of variables is polynomial in the number of agents. With these new variables we can still express the BIC and IIR constraints of the initial LP. %\grigorisnote{drop this sentence} We emphasize that even though the BIC constraints grow exponentially in the number of actions, we can use a separation oracle to check their feasibility in the same way as in the single-agent case. 
However, it is not yet clear how to check whether these variables correspond to an actual feasible mechanism. In the next section, we show how to design a separation oracle that checks the feasibility efficiently.

%This new LP consists of the variables $\hat{\pi}^i_{\omega, a}(\theta^i), \hat{t}^i(\theta^i), \hat{p}^i(\theta^i)$ which should be thought of as the expected values of the respective variables in the original LP that depend on the whole type profile.
\begin{figure}[h!]
\colorbox{MyGray}{
\begin{minipage}{1.02\textwidth} {
\noindent\textbf{Variables:}
\begin{itemize}
\item $\{\hat{\pi}^i_{\omega,j}(\theta^i)\}_{\omega\in \Omega, i\in [n], \theta^i\in \Theta^i, j\in[m]}$, denoting the reduced form of the mechanism.
\item $\{\hat{t}^i(\theta^i)\}_{i\in [n], \theta^i\in \Theta^i}$, denoting the interim prices.
\item $\{\hat{p}^i(\theta^i)\}_{i \in [n], \theta^i \in \Theta^i}$, denoting the allocation probabilities of the experiment
\item $\{\hat{z}^i_j(\theta^i, \tilde{\theta^i})\}_{i \in [n], j \in [m], \theta^i, \tilde{\theta^i} \in \Theta^i}$, helper variables. $\hat{z}^i_j(\theta^i, \tilde{\theta^i})$ represents an upper bound of the conditional expected utility of signal $s_j$ for type $\theta^i$.
\end{itemize}
\textbf{Linear Program:}
\begin{equation*}
\begin{array}{ll@{}ll}
\text{max}  &  \displaystyle\sum\limits_{i\in[n]} \displaystyle\sum\limits_{\theta^i\in \Theta^i}F^i(\theta^i) \hat{t}^i(\theta^i)&\\
\text{subject to}
  %  &             \displaystyle\sum\limits_{\omega}\theta^i_{\omega} \hat{\pi}^i_{\omega, a}(\theta^i)u_{\omega, a} \geq 
                    %\displaystyle\sum\limits_{\omega}\theta^i_{\omega} \hat{\pi}^i_{\omega, a}(\theta^i)u_{\omega, a'},  &\forall i, \forall \theta^i,  \forall a, \forall a'  ~~~(\text{Obedience})\\
                    
           % &\displaystyle\sum\limits_{\omega \in \Omega} \theta^i_{\omega} \hat{\pi}^i_{\omega, j}(\theta^i)u^i_{\omega, a_j} \geq  \displaystyle\sum\limits_{\omega \in \Omega} \theta^i_{\omega} \hat{\pi}^i_{\omega, j}(\theta^i)u^i_{\omega, a_k}, & \forall i, \forall \theta^i, \forall j, k ~~~ (\text{OB})\\
                &   \displaystyle\sum\limits_{j \in [m]}\displaystyle\sum\limits_{\omega \in \Omega} \theta^i_{\omega} \hat{\pi}^i_{\omega, j}(\theta^i)u^i_{\omega, a_j}+(1-\hat{p}^i(\theta^i))u^i(\theta^i) - \hat{t}^i(\theta^i) \geq\\
                
                 &~~~~~~~\displaystyle\sum\limits_{j\in[m]} \hat{z}^i_j(\theta^i, \tilde{\theta^i})+(1-\hat{p}^i(\tilde{\theta^i}))u^i(\theta^i) - \hat{t}^i(\tilde{\theta^{i}}) ,~~~ 
                &  
                \forall i,  \forall \theta^i , \tilde{\theta}^i  ~(\text{BIC})\\
                
                & \hat{z}^i_j(\theta^i, \tilde{\theta^i}) \geq \displaystyle\sum\limits_{\omega \in \Omega} \theta^i_\omega \hat{\pi}^i_{\omega, j}(\tilde{\theta^i}) u^i_{\omega, a_k} & \forall i, \forall \theta^i, \tilde{\theta^i}, \forall j, k\\
                
                    &   \displaystyle\sum\limits_{j \in [m]
                    }\displaystyle\sum\limits_{\omega \in \Omega} \theta^i_{\omega} \hat{\pi}^i_{\omega, j}(\theta^i)u^i_{\omega, a_j}+(1-\hat{p}^i(\theta^i))u^i(\theta^i) - \hat{t}^i(\theta^i) \geq u^i(\theta^i),
               ~~ &  \forall i, \forall \theta^i ~(\text{IIR})\\  
                &\hat{p}^i(\theta^i)=\displaystyle\sum\limits_{j\in[m]} \hat{\pi}^i_{\omega,j}(\theta^i) & \forall i, \forall \omega, \forall \theta^i\\
                
               & \{\hat{\pi}^i_{\omega,j}(\theta^i)\}_{\omega\in \Omega, i\in [n], \theta^i\in \Theta^i, j\in[m]}\in \PP(F)& 
                (\text{Feasibility})
\end{array}
\end{equation*}}
\end{minipage}} \caption{A linear program to find the reduced form of the revenue-optimal mechanism in the multi-agent setting.}\label{fig:alg multi-agent}
\end{figure}

\subsection{Feasibility of Reduced Forms}\label{sec:feasibility of reduced form}
To design a separation oracle for the set $\PP(F)$, we invoke the equivalence between Optimization and Separation in Linear Programming~\cite{GrotschelLS81,KarpP82}, which states that being able to optimize any linear function over a convex set $P$ is equivalent to having a separation oracle for $P$. It is a well-known fact that given a separation oracle for $P$ one can optimize any linear function using the ellipsoid method. Interestingly, the reverse is also true. If there is an algorithm to optimize any linear function over $P$, one can construct a separation oracle for $P$ using the ellipsoid method. We state a strengthened version of the equivalence due to Cai et al.~\cite{CaiDW13a}. The reason that we need to be able to decompose a feasible point into corners of the polytope is that we eventually need to able to implement the reduced forms as a feasible mechanism. We elaborate more on this later.

\begin{theorem}\label{thm:equivalence}(Adapted from Theorem H.1 of~\cite{CaiDW13a})
Let $P$ be a $d$-dimensional closed convex region, and let $\mathcal{A}$ be any polynomial-time algorithm that takes any direction ${w}\in \mathbb{R}^{d}$ as input and outputs the extreme point $\mathcal{A}({w}) \in P$ in direction ${w}$ such that $\mathcal{A}({w})\cdot {w}\geq \cdot \max_{{x}\in P} {x}\cdot{w}$ . Then we can design a polynomial time separation oracle $SO$ for $P$ such that, whenever $SO({x}) =$ ``yes'', the execution of $SO$ explicitly finds directions ${w}_1,\ldots,{w}_k$ such that ${x}$ lies in the convex hull of $\{\mathcal{A}({w}_1),\ldots, \mathcal{A}({w}_k)\}$.

%Let $P$ be any $d$-dimensional closed convex region and let $\mathcal{A}$ be an algorithm that optimizes linear functions over $P$. That is, $\mathcal{A}$ takes as input a $d$-dimensional vector $\vec{w}$ and outputs $\vec{x} \in \argmax_{\vec{x} \in P} \vec{x} \cdot \vec{w}$. Then there exists a separation oracle $SO$ for $P$ such that the running time of the separation oracle on a $b$-bit input is $\rt_{SO}(b) = \poly(d, b, \rt_{\mathcal{A}}(\poly(d, b))$, where $\rt_{\mathcal{A}}(x)$ is the running time for algorithm $\mathcal{A}$ on an $x$-bit input.
\end{theorem}

To apply this equivalence, we need to show how we can optimize a linear function over the set of feasible reduced-form variables. Recall that we use $\Pi^i(\theta) = \left( \pi^i_{\omega, j}(\theta) \right)_{\omega \in \Omega, j \in [m]}$ and $\hat{\Pi}^i(\theta^i) = \left( \hat{\pi}^i_{\omega, j}(\theta^i) \right)_{\omega \in \Omega, j \in [m]}$ to denote the ex-post signaling scheme and its reduced form. We will treat $\Pi^i(\theta)$ and $\hat{\Pi}^i(\theta^i)$ as $m|\Omega|$-dimensional vectors. The following maximization problem plays a crucial role in our approach.

\begin{definition}
Consider any type profile $\theta$. Let  $\{X^i(\theta^i)\}_{i \in [n],\theta_i\in \Theta_i}$ be a collection of $m|\Omega|$-dimensional vectors. We define a Virtual Payoff Maximizer (VPM) w.r.t. these weight vectors VPM$(\{X^i(\theta^i)\}_{i \in [n],\theta_i\in\Theta_i})$ to be the ex-post signaling scheme $\Pi(\theta)$ that maximizes the following quantity $ \sum_{i} \Pi^i(\theta^i)\cdot X^i(\theta^i)$
for every type profile $\theta$. The corresponding reduced form $\hat{\Pi}(\theta)$ is called rVPM$(\{X^i(\theta^i)\}_{i \in [n],\theta^i\in\Theta^i})$. 
In order to ensure that the maximizer is unique, we break ties lexicographically.
\end{definition}

When there is no confusion, we also write $VPM({w}), rVPM({w})$ as the maximizers for the weight vector {${w}$}.

\begin{lemma}\label{lem:maximizer multi-agent}
    Given an arbitrary collection of weights $\{X^i(\theta^i)\}_{i \in [n],\theta^i\in \Theta^i}$ we can find the exact optimal solution of  \\$\max_{\hat{\Pi}\in \PP(F)} \sum_{i\in[n]}\sum_{\theta^i\in \Theta^i} \hat{\Pi}^i(\theta^i) \cdot X^i(\theta^i)$ in time $O\left(m|\Omega|\left(\sum_{i \in [n]} |\Theta^i|\right)+\left(\sum_{i \in [n]} |\Theta_i|\right)^2\right)$.
\end{lemma}

\begin{prevproof}{Lemma}{lem:maximizer multi-agent}
We first rewrite the maximization problem

\begin{align*}
  \max_{\hat{\Pi}\in \PP(F)} \sum_{i\in[n]}\sum_{\theta^i\in \Theta^i} \hat{\Pi}^i(\theta^i) \cdot X^i(\theta^i) =
    \max_{\Pi} \sum_{i, \theta^i, \theta^{-i}} F^{-i}(\theta^{-i}) \Pi^i(\theta^i, \theta^{-i}) \cdot X^i(\theta^i) = \\
    \max_{\Pi} \sum_{i, \theta} F(\theta) \Pi^i(\theta) \cdot \frac{X^i(\theta^i)}{F^i(\theta^i)} =
    \max_{\Pi} \sum_{\theta} F(\theta) \sum_i \Pi^i(\theta) \cdot \tilde{X}^i(\theta^i),
\end{align*}
where $\tilde{X}^i(\theta^i) = \frac{X^i(\theta^i)}{F^i(\theta^i)}$.

    Let $\theta$ be a type profile. We now characterize the solution of $\max_{\Pi(\theta)} \sum_i \Pi^i(\theta) \cdot \tilde{X}^i(\theta^i)$. %in time $O(nm|\Omega|)$.
    If we allocate the experiment to buyer $i$, the maximum value we can derive is $v^i(\theta^i) = \sum_\omega \max_a 
    \tilde{X}^i_{\omega, a}(\theta^i)$. Clearly, the optimal solution of the linear function above is to always allocate the experiment to the buyer with the largest $v^i(\theta^i)$.
    
    The ex-post signaling scheme that maximizes $\sum_{\theta} F(\theta) \sum_i \Pi^i(\theta) \cdot \tilde{X}^i(\theta^i)$ is the one that always allocates the experiment to the buyer with the largest $v^i(\theta^i)$ for every type profile $\theta$. To solve $  \max_{\hat{\Pi}\in \PP(F)} \sum_{i\in[n]}\sum_{\theta^i\in \Theta^i} \hat{\Pi}^i(\theta^i) \cdot X^i(\theta^i)$, we only need to calculate the reduced form of this ex-post signaling scheme and we denote it using $\hat{\Pi}_*$
    
   % Assume that the optimal solution allocates mass $0< c_i \leq 1$ to agent $i$, i.e. \sum_{j\in[m]} $\Pi^i_{\omega,j} (\theta)$ is $c_i$ for each $\omega$. Fix a state $\omega$ of $\tilde{X}^i(\theta^i)$ and let $a = \argmax_{a \in A} x_{\omega, a}^i(\theta^i)$ be the maximum element of this row. Since every $\pi^i_{\omega_, a}(\theta) \geq 0$, the optimal choice for this row is to set $\pi^i_{\omega, a}(\theta) = c_i, \pi^i_{\omega, a'} = 0, a' \neq a$. Hence, the maximum contribution of agent $i$ to the solution is $c_i \sum_\omega \max_a x^i_{\omega, a}(\theta^i)$. Let $v^i(\theta^i) = \sum_\omega \max_a x^i_{\omega, a}(\theta^i)$. We see that if we want to maximize the objective, we better allocate all the mass to the agent that has the highest $v^i(\theta^i)$. In case we have ties, we break them lexicographically.

%So far we have seen that once we fix the type profile, it is easy to find the allocation that maximizes the objective. However, this is not enough to create the separation oracle since there are exponentially many type profiles. The following lemma shows how we can overcome this obstacle.

%\begin{lemma}\label{lem: multi-private, exact max}
 
 %   Given an arbitrary linear function $\{X^i(\theta^i)\}_{i \in [n]}$ we can find the exact optimal solution of $\max_{\Pi} \sum_{\theta} F(\theta) \sum_i \Pi^i(\theta) \cdot X^i(\theta^i)$, where the maximization is taken over the set of all feasible allocation rules, in time $O\left(\left(\sum_{i \in [n]} |\Theta_i|\right)^2\right)$.\end{lemma}

We first compute $v^i(\theta^i) = \sum_\omega \max_a x^i_{\omega, a}(\theta^i)$ for every buyer $i$ and every type $\theta^i$. This step takes time $O\left(m|\Omega|\left(\sum_{i \in [n]} |\Theta^i|\right)\right)$, and there are $\sum_{i \in [n]} |\Theta^i|$ different such values. Next, for each buyer, we sort $v^i(\theta^i)$. This step takes time $O\left(\sum_{i \in [n]} |\Theta^i|\log |\Theta^i|\right)$.  To compute $\hat{\Pi}_*^i(\theta^i)$, we only need to calculate the probability of the event that over the random draws of $\theta_{-i}$, there exists another buyer $\ell\neq i$ either $v_\ell(\theta^\ell)>v^i(\theta^i)$ or $\ell<i$ and $v_\ell(\theta^\ell)=v^i(\theta_i)$. This probability can be computed in time $O\left(\sum_{i \in [n]} |\Theta^i|\right)$ for each buyer $i$ and type $\theta^i$. Hence, in total we can optimize the linear function in time $O\left(m|\Omega|\left(\sum_{i \in [n]} |\Theta^i|\right)+\left(\sum_{i \in [n]} |\Theta_i|\right)^2\right)$.

\end{prevproof}

Combining Theorem~\ref{thm:equivalence} and Lemma~\ref{lem:maximizer multi-agent}, we have a polynomial time algorithm to solve the LP in Figure~\ref{fig:alg multi-agent}, but we still need to turn the reduced form into an ex-post signaling scheme. We again use an idea from computing the optimal multi-item auctions, that is, first decomposing the optimal reduced form into a distribution over extreme points of $\PP(F)$, then implementing all the extreme points that appear in the distribution using a VPM ex-post signaling scheme.

\begin{theorem}\label{thm:mutli-agent}
We design an algorithm to compute the revenue-optimal mechanism in time $\poly\left(n,m,|\Omega|,\sum_{i\in[n]} |\Theta^i|\right)$. Moreover, the mechanism can be implemented as a distribution over $m|\Omega| \left(\sum_{i\in[n]} |\Theta^i|\right)+1$ VPM ex-post signaling schemes.
\end{theorem}

\begin{prevproof}{Theorem}{thm:mutli-agent}
Combining Theorem~\ref{thm:equivalence} and Lemma~\ref{lem:maximizer multi-agent}, we have a polynomial time algorithm to solve the LP in Figure~\ref{fig:alg multi-agent}. Let $\Pi_*=\left\{\hat{\Pi}_*^{i}(\theta^i)\right\}_{i\in[n],\theta^i\in\Theta^i}$ be the optimal reduced form and $\left\{\hat{t}_*^i(\theta^i)\right\}_{i\in[n],\theta^i\in\Theta^i}$ be the optimal interim prices. Since the SO must return ``yes'' on $\Pi_*$, as guaranteed by Theorem~\ref{thm:equivalence}, the SO also finds a collection of directions $w_1,\ldots, w_k$ so that $\Pi_*$ lies in the convex hull of $rVPM(w_1),\ldots, rVPM(w_k)$. Due to the Carath\'{e}odory's theorem, we can decompose $\Pi_*$ into a distribution $D$ over at most $m|\Omega| \left(\sum_{i\in[n]} |\Theta^i|\right)+1$ of the above rVPMs. Moreover, we can use a LP to find this distribution in polynomial time. To implement $\Pi_*$, we first sample a $rVPM(w)$ from the distribution $D$, and implement the corresponding ex-post signaling scheme $VPM(w)$. Finally, for each buyer $i$, we charge him $\hat{t}^i(\theta^i)$ if he reports $\theta^i$, so the mechanism is BIC and IIR, and we do not reveal extra information through the prices.
\end{prevproof}

\section{Further Extensions and Future Directions}\label{sec:further extensions}
In this section, we discuss further generalizations
    of the model by Bergemann et al. \cite{bergemann2018design} {and future research directions that we believe are interesting to pursue}. 
    
    \subsection{Extensions of the Original Model}
    \paragraph*{Enlarged Buyer Type:} Recall that in the original model the only private information of the buyer is his private belief of the underlying state, which is realized at the beginning of the interaction with the seller. Importantly, the payoffs are public knowledge and remain the same across different buyers. 
    A natural generalization one can consider is to allow the buyer to draw not only his prior belief $\theta$, but also his payoff function $u: \Omega \times A \rightarrow [0,1]$ from some distribution. 
    
    To be more specific, we consider the setting where a buyer's type $\rho=(\theta, u)$ is drawn from some distribution $F$ at the beginning of the interaction between the buyer and the seller, and $\rho$ is private to the buyer. As in the original model, we assume that the seller has access to this distribution. We remark that all of our positive results from Section~\ref{sec:single-agent} and Section~\ref{sec:multi-agent}, except for Theorem~\ref{thm:arbitrary theta}, hold in this extended model as well. The only difference in our constructions is that instead of indexing the variables by $\theta$ we now index them by $\rho$.
    
\paragraph*{Misspecified Model:}    Another generalization we consider in the single-agent setting is the \textit{misspecified} model. In this model the seller has access to some type distribution $\Tilde{F}$ which is within $\varepsilon$ in TV-distance with the real type distribution $F$. Moreover, the seller has access to a type space $\tilde{\Theta}$ with the following two properties: $|\tilde{\Theta}| = |\Theta|$, for all $\Tilde{\theta} \in \tilde{\Theta}$ there is some $\theta \in \Theta$ for which $d_{TV}(\Tilde{\theta}, \theta) \leq \varepsilon$. Then, the menu that the seller designs for the misspecified distributions can be modified so that it guarantees only a negligible revenue loss when it is evaluated in the true setting. Lemma~\ref{lem:misspecified setting} formalizes this claim.

\begin{lemma}\label{lem:misspecified setting}
    Let $\Tilde{F}, F$ be the distributions of the types that the seller has access to and the true distribution of the types, respectively. Let also $\{\Tilde{\theta}_i\}_{i \in [k]}, \{\theta_i\}_{i \in [k]}$ be the types that the seller has access to and the true types, respectively. Assume that $d_{TV}(\Tilde{F}, F) \leq \varepsilon_1, d_{TV}(\Tilde{\theta_i}, \theta_i) \leq \varepsilon_2, \forall i \in [k]$. We also let $\widetilde{\MM} = \{\tilde{E}(\tilde{\theta}_i), \tilde{t}(\tilde{\theta}_i)\}_{i \in [k]}$ be an IC, IR menu that has revenue $\rev(\widetilde{\MM})$ under the misspecified distributions and uses at most $|S|$ signals. Then, we can compute a set of prices $\{t(\tilde{\theta}_i)\}_{i \in [k]} $ so that $\MM = \{\tilde{E}(\theta_i), t(\theta_i)\}_{i \in [k]}$ is IR, IC and has $\rev(\MM) \geq \rev(\widetilde{\MM}) - O\left(\varepsilon_1 + \sqrt{|\Omega|\varepsilon_2} \right)$ under the true distributions.
\end{lemma}

\begin{prevproof}{Lemma}{lem:misspecified setting}
        We first show that $\widetilde{\MM}$ violates the IC, IR constraints by at most $2|\Omega|\varepsilon_2$. Consider a type $\Tilde{\theta}$ who buys $E$, but the true type $\theta$ prefers $E'$ over $E$. Then, by Lemma~\ref{lem:change offers theta}, $V_{\theta}(E) - \tilde{t}(E) \geq V_{\theta}(E') - \tilde{t}(E') - 2|\Omega|\varepsilon_2$. Hence, by Lemma~\ref{lem: eps-bic to bic} we know that we can compute a new set of prices in time $O(|\widetilde{\MM}|)$ losing at most $O(\sqrt{|\Omega|\varepsilon_2})$ revenue. Moreover, since the seller's distribution over the types is also misspecified we have
        
        \begin{align*}
            \sum_\theta \left(\tilde{F}(\theta) - F(\theta)\right)t(\theta) \leq 2\varepsilon_1 \implies 
            \sum_\theta F(\theta)t(\theta) \geq \sum_\theta \tilde{F}(\theta) t(\theta) - 2\varepsilon_1 \geq \\ 
            \sum_{\tilde{\theta}} \tilde{F}(\tilde{\theta}) \tilde{t}(\tilde{\theta}) - 2\varepsilon_1 - O(\sqrt{|\Omega|\varepsilon_2})
        \end{align*}
\end{prevproof}

Note that Lemma~\ref{lem:misspecified setting} allows us to generalize our results and obtain approximately-optimal menus when we only have black-box access to the distribution of the types. That is, we take enough samples to learn the distribution within Total Variation distance $\varepsilon$ and then apply Lemma~\ref{lem:misspecified setting}.

\subsection{Future Directions}
We believe that the design of Information Markets is a very important problem that has not received sufficient attention by the Theory of Computation community. There are many interesting questions waiting be addressed.

\begin{enumerate}
    \item In the single-buyer setting where we only have access to the action space via a BR oracle the running time of our algorithms is exponential in the number of states. An immediate question to ask is whether we can get an FPTAS or even a PTAS that has a better dependence on the number of states.
    
    \item In the multi-agent setting, we consider the case in which the seller does not reveal the signaling scheme that she uses to send a signal to the winner. An interesting question is whether we can have efficient algorithms in the setting where the seller reveals the signaling scheme to the buyer.
    
    \item Currently, in the multi-agent setting we assume that the ex-post utility of each buyer depends only on the state of the world and the action he takes. Is the problem of designing the optimal mechanism when the ex-post utilities also depend on the actions of the other buyers tractable?
    
    %\item Our paper focuses on the setting where there is only one seller who knows the state of the world. However, we believe that a more realistic scenario is the one in which there are many sellers who possess information of a different quality. How does the presence of multiple sellers affect the tractability of designing the revenue-optimal mechanism for each one of them?
\end{enumerate}

 %\newpage
 %\bibliographystyle{plain}
 \bibliography{Yang.bib}
 
 %\newpage
%\appendix
%\input{appendix}

\end{document}